\documentclass[11pt,a4paper]{article}
\usepackage{amssymb}
\usepackage{amsmath}
\usepackage{hyperref}
\usepackage[page,title]{appendix}

\renewcommand{\theequation}{\thesection.\arabic{equation}}
\input{tcilatex}
\begin{document}

\title{Local geometry of the $G_{2}$ moduli space}
\author{Sergey Grigorian \\
DAMTP\\
Centre for Mathematical Sciences\\
Wilberforce Road\\
Cambridge CB3 0WA\\
United Kingdom \and Shing-Tung Yau \\
Department of Mathematics\\
Harvard University\\
Cambridge, MA 02138\\
USA}
\maketitle

\begin{abstract}
We consider deformations of torsion-free $G_{2}$ structures, defined by the $%
G_{2}$-invariant $3$-form $\varphi $ and compute the expansion of $\ast
\varphi $ to fourth order in the deformations of $\varphi $. By considering $%
M$-theory compactified on a $G_{2}$ manifold, the $G_{2}$ moduli space is
naturally complexified, and we get a K\"{a}hler metric on it. Using the
expansion of $\ast \varphi $ we work out the full curvature of this metric
and relate it to the Yukawa coupling.
\end{abstract}

\section{Introduction}

\setcounter{equation}{0}One of the possible approaches to $M$-theory is to
consider compactifications of the $11$-dimensional spacetimes of the form $%
M_{4}\times X$ where $M_{4}$ is the $4$-dimensional Minkowski space and $X$
is a $7$-dimensional manifold. If $X$ is a compact manifold with $G_{2}$
holonomy, then this gives a vacuum solution of the low-energy effective
theory, and moreover, since $X$ has one covariantly constant spinor, the
resulting theory in $4$ dimensions has $N=1$ supersymmetry. The physical
content of the $4$-dimensional theory is given by the moduli of $G_{2}$
holonomy manifolds. Such a compactification of $M$-theory is in many ways
analogous to Calabi-Yau compactifications in String Theory, where much
progress has been made through the study of the Calabi-Yau moduli spaces. In
particular, as it was shown in \cite{Candelas:1990pi} and \cite%
{Strominger:1990pd}, the moduli space of complex structures and the
complexified moduli space of K\"{a}hler structures are both in fact, K\"{a}%
hler manifolds. Moreover, both have a \emph{special geometry} - that is,
both have a line bundle whose first Chern class coincides with the K\"{a}%
hler class. However until recently, the structure of the moduli space of $%
G_{2}$ holonomy manifolds has not been studied in that much detail.
Generally, it turned out that the study of $G_{2}$ manifolds is quite
difficult. Firstly, unlike in the Calabi-Yau case \cite{CalabiYau}, there is
no general theorem for existence of $G_{2}$ manifolds. Although there are
constructions of compact $G_{2}$ manifolds such as those that can be found
in \cite{Joycebook} and \cite{Kovalev:2001zr}, they are not explicit (a
non-compact construction was also given in \cite{Gibbons:1989er}). Another
difficulty is that the $G_{2}$-invariant $3$-form which defines the $G_{2}$%
-structure and the metric corresponding to it are related in a non-linear
fashion. This makes the study of $G_{2}$ manifolds more difficult from a
computational point of view.

We first start with an overview of $G_{2}$ structures in section 2, where we
state the basic facts about $G_{2}$ manifolds and set up the notation. A $%
G_{2}$-structure is defined by a $G_{2}$-invariant $3$-form $\varphi $, and
in section 3 we review some of the computational properties of $\varphi $
and its Hodge dual $\ast \varphi $, which we will need later on. Since one
of our main motivation to study $G_{2}$ manifolds comes from physics, in
section 4, we review the role of $G_{2}$ manifolds in $M$-theory, and in
particular we consider the Kaluza-Klein compactification of the effective $M$%
-theory low-energy action on a $G_{2}$ manifold. It turns that in the
reduced action, the moduli of the $M$-theory $3$-form $C_{mnp}$ and the $%
G_{2}$ moduli naturally combine, to effectively give a complexification of
the $G_{2}$ moduli space. Moreover, the metric on this complexified space
turns out to be K\"{a}hler, and the K\"{a}hler potential is essentially the
logarithm of the volume of the $G_{2}$ manifold.

The aim of this paper is to gain more information about the geometry of the
moduli space, and so the aim is to compute the curvature of this K\"{a}hler
metric. This involves calculation of the fourth derivative of the K\"{a}hler
potential. The method which we use for this requires us to know the
expansion of $\ast \varphi $ to third order in the deformations of $\varphi $%
. So in section 5, we in fact explicitly give the expansion of $\ast \varphi 
$ to fourth order in the deformations of $\varphi $. Previously, only the
full expansion to first order was known \cite{Joycebook}, and only partially
to second order \cite{bryant-2003}. However, there are approaches to
calculating higher derivatives of the K\"{a}hler potential without
explicitly computing an expansion of $\ast \varphi $ - for example the third
derivative has been computed by de Boer et al in \cite{deBoer:2005pt} and by
Karigiannis and Leung in \cite{karigiannis-2007a}.

Finally, in section 6, we use our expansion of $\ast \varphi $ from section
5 to calculate the full curvature of the $G_{2}$ moduli space, and then the
Ricci curvature as well. As it has already been noted in \cite{deBoer:2005pt}
and \cite{karigiannis-2007a}, the third derivative of the K\"{a}hler can be
interpreted as a Yukawa coupling, and it bears a great resemblance to the
Yukawa coupling encountered in the study of Calabi-Yau moduli spaces. At the
end of section 6 we consider look at some properties of covariant
derivatives on the moduli space.

\textbf{Acknowledgements.} The first author would like to thank Spiro
Karigiannis and Alexei Kovalev for useful discussions, and would also like
to thank UC\ Irvine and Harvard University, where much of this work has been
completed, for hospitality. The research of the first author is funded by
EPSRC.

\section{Overview of $G_{2}$ structures}

We will first review the basics of $G_{2}$ structures on smooth manifolds.
The main references for this section are \cite{Joycebook},\cite{bryant-2003}
and \cite{karigiannis-2005-57}.

The $14$-dimensional Lie group $G_{2}$ can be defined as a subgroup of $%
GL\left( 7,\mathbb{R}\right) $ in the following way. Suppose $%
x^{1},...,x^{7} $ are coordinates on $\mathbb{R}^{7}$ and let $%
e^{ijk}=dx^{i}\wedge dx^{j}\wedge dx^{k}$. Then define $\varphi _{0}$ to be
the $3$-form on $\mathbb{R}^{7}$ given by 
\begin{equation}
\varphi _{0}=e^{123}+e^{145}+e^{167}+e^{246}-e^{257}-e^{347}-e^{356}.
\label{phi0def}
\end{equation}%
Then $G_{2}$ is defined as the subgroup of $GL\left( 7,\mathbb{R}\right) $
which preserves $\varphi _{0}$. Moreover, it also fixes the standard
Euclidean metric 
\begin{equation}
g_{0}=\left( dx^{1}\right) ^{2}+...+\left( dx^{7}\right) ^{2}  \label{g0def}
\end{equation}
on $\mathbb{R}^{7}$ and the $4$-form $\ast \varphi _{0}$ which is the
corresponding Hodge dual of $\varphi _{0}$:%
\begin{equation}
\ast \varphi
_{0}=e^{4567}+e^{2367}+e^{2345}+e^{1357}-e^{1346}-e^{1256}-e^{1247}.
\label{sphi0def}
\end{equation}

Now suppose $X$ is a smooth, oriented $7$-dimensional manifold. A $G_{2}$
structure $Q$ on $X$ is a principal subbundle of the frame bundle $F$, with
fibre $G_{2}$. However we can also uniquely define $Q$ via $3$-forms on $X.$
Define a $3$-form $\varphi $ to be \emph{positive }if we locally can choose
coordinates such that $\varphi $ is written in the form (\ref{phi0def}) -
that is for every $p\in X$ there is an isomorphism between $T_{p}X$ and $%
\mathbb{R}^{7}$ such that $\left. \varphi \right\vert _{p}=\varphi _{0}$.
Using this isomorphism, to each positive $\varphi $ we can associate a
metric $g$ and a Hodge dual $\ast \varphi $ which are identified with $g_{0}$
and $\ast \varphi _{0}$ under this isomorphism. and the associated metric is
written (\ref{g0def}). It is shown in \cite{Joycebook} that there is a $1-1$
correspondence between positive $3$-forms $\varphi $ and $G_{2}$ structures $%
Q$ on $X$.

So given a positive $3$-form $\varphi $ on $X$, it is possible to define a
metric $g$ associated to $\varphi $ and this metric then defines the Hodge
star, which in turn gives the $4$-form $\ast \varphi $. Thus although $\ast
\varphi $ looks linear in $\varphi $, it actually is not, so sometimes we
will write $\psi =\ast \varphi $ to emphasize that the relation between $%
\varphi $ and $\ast \varphi $ is very non-trivial.

In general, any $G$-structure on a manifold $X$ induces a splitting of
bundles of $p$-forms into subbundles corresponding to irreducible
representations of $G$. The same is of course true for $G_{2}\,$-structure.
From \cite{Joycebook} we have the following decomposition of the spaces of $%
p $-forms $\Lambda ^{p}$: 
\begin{subequations}
\label{formdecompose}
\begin{eqnarray}
\Lambda ^{1} &=&\Lambda _{7}^{1}  \label{l1decom} \\
\Lambda ^{2} &=&\Lambda _{7}^{2}\oplus \Lambda _{14}^{2}  \label{l2decom} \\
\Lambda ^{3} &=&\Lambda _{1}^{3}\oplus \Lambda _{7}^{3}\oplus \Lambda
_{27}^{3}  \label{l3decom} \\
\Lambda ^{4} &=&\Lambda _{1}^{4}\oplus \Lambda _{7}^{4}\oplus \Lambda
_{27}^{4}  \label{l4decom} \\
\Lambda ^{5} &=&\Lambda _{7}^{5}\oplus \Lambda _{14}^{5}  \label{l5decom} \\
\Lambda ^{6} &=&\Lambda _{7}^{6}  \label{l6decom}
\end{eqnarray}

Here each $\Lambda _{k}^{p}$ corresponds to the $k\,$-dimensional
irreducible representation of $G_{2}$. Moreover, for each $k$ and $p$, $%
\Lambda _{k}^{p}$ and $\Lambda _{k}^{7-p}$ are isomorphic to each other via
Hodge duality, and also $\Lambda _{7}^{p}$ are isomorphic to each other for $%
n=1,2,...,6$. Note that $\varphi $ and $\ast \varphi $ are $G_{2}$%
-invariant, so they generate the $1$-dimensional sectors $\Lambda _{1}^{3}$
and $\Lambda _{1}^{4}$, respectively.

Define the standard inner product on $\Lambda ^{p},$ so that for $p\,$-forms 
$\alpha $ and $\beta $, 
\end{subequations}
\begin{equation}
\left\langle \alpha ,\beta \right\rangle =\frac{1}{p!}\alpha
_{a_{1}...a_{p}}\beta ^{a_{1}...a_{p}}\text{.}  \label{forminp}
\end{equation}%
This is related to the Hodge star, since 
\begin{equation}
\alpha \wedge \ast \beta =\left\langle \alpha ,\beta \right\rangle \mathrm{%
vol}  \label{hodgedef}
\end{equation}%
where $\mathrm{vol}$ is the invariant volume form given locally by 
\begin{equation}
\mathrm{vol}=\sqrt{\det g}dx^{1}\wedge ...\wedge dx^{7}.  \label{voldef}
\end{equation}%
Then it turns out that the decompositions (\ref{formdecompose}) are
orthogonal with respect to (\ref{forminp}). This will be seen easily when we
consider these decompositions in more detail in the next section.

As we already know, the metric $g$ on a manifold with $G_{2}$ structure is
determined by the invariant $3$-form $\varphi $. It is in fact possible to
write down an explicit relationship between $\varphi $ and $g$. Let $u$ and $%
v$ be vector fields on $X$. Then 
\begin{equation}
\left\langle u,v\right\rangle \mathrm{vol}=\frac{1}{6}\left( u\lrcorner
\varphi \right) \wedge \left( v\lrcorner \varphi \right) \wedge \varphi .
\label{metricdef}
\end{equation}%
Here $\lrcorner $ denotes interior multiplication, so that 
\begin{equation}
\left( u\lrcorner \varphi \right) _{bc}=u^{a}\varphi _{abc}.
\label{intmultdef}
\end{equation}%
The definition (\ref{metricdef}) is rather indirect because $\mathrm{vol}$
depends on $g$ via (\ref{voldef}). To make more sense of it, rewrite in
components%
\begin{equation}
g_{ab}\sqrt{\det g}=\frac{1}{144}\varphi _{amn}\varphi _{bpq}\varphi _{rst}%
\hat{\varepsilon}^{mnpqrst}  \label{metriccomp1}
\end{equation}%
where $\hat{\varepsilon}^{mnpqrst}$ is the alternating symbol with $%
\varepsilon ^{12...7}=+1$. Define 
\begin{equation}
B_{ab}=\frac{1}{144}\varphi _{amn}\varphi _{bpq}\varphi _{rst}\hat{%
\varepsilon}^{mnpqrst}  \label{sabdef}
\end{equation}%
so that then, after taking the determinant of (\ref{metriccomp1}) we get 
\begin{equation}
g_{ab}=\left( \det B\right) ^{-\frac{1}{9}}B_{ab}.  \label{metricdefdirect}
\end{equation}%
This gives a direct definition, but because $\det s$ may be awkward to
compute, (\ref{metricdefdirect}) is not always the most practical
definition. For us, it will be more useful to take the trace of (\ref%
{metriccomp1}) with respect to $g$, which gives%
\begin{equation}
\sqrt{\det g}=\frac{1}{7}\func{Tr}B  \label{detgtrs}
\end{equation}%
and hence 
\begin{equation}
g_{ab}=\frac{7B_{ab}}{\func{Tr}B}.  \label{gabtrs1}
\end{equation}%
Although this is also an indirect definition, it is sometimes easier to
handle this expression.

There are in fact a total of 16 torsion classes of $G_{2}$ structures, each
of which places certain restrictions on $d\varphi $ or $d\ast \varphi $ \cite%
{FernandezGray}. One of the most important classes of manifolds with $G_{2}$
structure are manifolds with $G_{2}$ holonomy. The group $G_{2}$ appears as
one of two exceptional holonomy groups - the other one is $Spin\left(
7\right) $ for $8$-dimensional manifolds. The list of possible holonomy
groups is limited and they were fully classified by Berger \cite{Berger1955}%
. Specifically, if $\left( X,g\right) $ is a simply-connected Riemannian
manifold which is neither locally a product nor is symmetric, the only
possibilities are shown in the table below. 
\begin{equation*}
\begin{tabular}{lll}
\textbf{Dimension} & \textbf{Holonomy} & \textbf{Type of Manifold} \\ 
$2k$ & \thinspace $U\left( k\right) $ & K\"{a}hler \\ 
$2k$ & $SU\left( k\right) $ & Calabi-Yau \\ 
$4k$ & $Sp\left( k\right) $ & HyperK\"{a}hler \\ 
$4k$ & $Sp\left( k\right) Sp\left( 1\right) $ & Quaternionic \\ 
$7$ & $G_{2}$ & Exceptional \\ 
$8$ & $Spin\left( 7\right) $ & Exceptional%
\end{tabular}%
\end{equation*}

It turns out that the holonomy group $Hol\left( X,g\right) \subseteq G_{2}$
if and only if $X$ has a torsion-free $G_{2}$ structure \cite{Joycebook}. In
this case, the invariant $3$-form $\varphi $ satisfies%
\begin{equation}
d\varphi =d\ast \varphi =0  \label{torsionfreedef}
\end{equation}%
and equivalently, $\nabla \varphi =0$ where $\nabla $ is the Levi-Civita
connection of $g$. So in fact, in this case $\varphi $ is harmonic.
Moreover, if $Hol\left( X,g\right) \subseteq G_{2}$, then $X$ is Ricci-flat.

For a torsion-free $G_{2}$ structure, the decompositions (\ref{formdecompose}%
) carry over to de Rham cohomology \cite{Joycebook}, so that we have 
\begin{subequations}
\label{cohodecom}
\begin{eqnarray}
H^{2}\left( X,\mathbb{R}\right) &=&H_{7}^{2}\oplus H_{14}^{2} \\
H^{3}\left( X,\mathbb{R}\right) &=&H_{1}^{3}\oplus H_{7}^{3}\oplus H_{27}^{3}
\\
H^{4}\left( X,\mathbb{R}\right) &=&H_{1}^{4}\oplus H_{7}^{4}\oplus H_{27}^{4}
\\
H^{5}\left( X,\mathbb{R}\right) &=&H_{7}^{5}\oplus H_{14}^{5}
\end{eqnarray}%
Define the refined Betti numbers $b_{k}^{p}=\dim \left( H_{k}^{p}\right) $.
Clearly, $b_{1}^{3}=b_{1}^{4}=1$ and we also have $b_{1}=b_{7}^{k}$ for $%
k=1,...,6$. Moreover, it turns out that $b_{1}=0$ if and only if $Hol\left(
X,g\right) =G_{2}$. Therefore, in this case the $H_{7}^{k}$ component
vanishes in (\ref{cohodecom}).

An example of a construction of a manifold with a torsion-free $G_{2}$
structure is to consider $X=Y\times S^{1}$ where is a Calabi-Yau $3$-fold.
Define the metric and a $3$-form on $X$ as 
\end{subequations}
\begin{eqnarray}
g_{X} &=&d\theta ^{2}\times g_{Y}  \label{metCY} \\
\varphi &=&d\theta \wedge \omega +\func{Re}\Omega  \label{phiCY}
\end{eqnarray}%
where $\theta $ is the coordinate on $S^{1}$. This then defines a
torsion-free $G_{2}$ structure, with 
\begin{equation}
\ast \varphi =\frac{1}{2}\omega \wedge \omega -d\theta \wedge \func{Im}%
\Omega .  \label{psiCY}
\end{equation}%
However, the holonomy of $X$ in this case is $SU\left( 3\right) \subset
G_{2} $. From the K\"{u}nneth formula we get the following relations between
the refined Betti numbers of $X$ and the Hodge numbers of $Y$ 
\begin{eqnarray*}
b_{7}^{k} &=&1\ \ \ \text{for }k=1,...,6 \\
b_{14}^{k} &=&h^{1,1}-1\ \ \text{for }k=2,5 \\
b_{27}^{k} &=&h^{1,1}+2h^{2,1}\ \text{\ for }k=3,4.
\end{eqnarray*}

\section{Properties of $\protect\varphi $}

The invariant $3$-form $\varphi $ which defines a $G_{2}$ structure on the
manifold $X$ has a number of useful and interesting properties. In
particular, contractions of $\varphi $ and $\psi =\ast \varphi $ are very
useful in computations. From \cite{bryant-2003}, \cite{karigiannis-2007} and 
\cite{House:2004pm}, we have 
\begin{eqnarray}
\varphi _{abc}\varphi _{mn}^{\ \ \ c} &=&g_{am}g_{bn}-g_{an}g_{bm}+\psi
_{abmn}  \label{phiphi1} \\
\varphi _{abc}\psi _{mnp}^{\ \ \ \ \ \ c} &=&3\left( g_{a[m}\varphi
_{np]b}-g_{b[m}\varphi _{np]a}\right)  \label{phipsi}
\end{eqnarray}%
Essentially, these identities can be derived straight from the definitions
of $\varphi $ and $\psi =\ast \varphi $ in flat space - (\ref{phi0def}) and (%
\ref{sphi0def}) respectively. For more details, please refer to \cite%
{bryant-2003} and \cite{karigiannis-2007}. Note that we are using a
different convention to \cite{karigiannis-2007}, and hence some of the signs
are different.

Consider the product $\psi _{abcd}\psi ^{mnpq}$. Expanding $\psi $ as the
Hodge star of $\varphi $ and then using the usual identity for a product of
Levi-Civita tensors and then applying (\ref{phiphi1}) gives 
\begin{equation}
\psi _{abcd}\psi ^{mnpq}=24\delta _{a}^{[m}\delta _{b}^{n}\delta
_{c}^{p}\delta _{d}^{q]}+72\psi _{\lbrack ab}^{\ \ \ [mn}\delta
_{c}^{p}\delta _{d]}^{q]}-16\varphi _{\lbrack abc}\varphi ^{\lbrack
mnp}\delta _{d]}^{q]}  \label{psipsi0}
\end{equation}%
Contracting over $d$ and $q$ gives%
\begin{equation}
\psi _{abcd}\psi ^{mnpd}=6\delta _{a}^{[m}\delta _{b}^{n}\delta
_{c}^{p]}+9\psi _{\lbrack ab}^{\ \ \ \ [mn}\delta _{c]}^{p]}-\varphi
_{abc}\varphi ^{mnp}  \label{psipsi1}
\end{equation}%
which agrees with the expression given in \cite{House:2004pm}. Of course the
above relations can be further contracted to obtain 
\begin{eqnarray}
\varphi _{abc}\varphi _{m}^{\ \ \ bc} &=&6g_{am}  \label{phiphi2} \\
\varphi _{abc}\psi _{mn}^{\ \ \ \ \ \ bc} &=&4\varphi _{amn}  \label{phipsi2}
\\
\psi _{abcd}\psi _{mn}^{\ \ \ \ \ cd} &=&4g_{am}g_{bn}-4g_{an}g_{bm}+2\psi
_{abmn}.  \label{psipsi2}
\end{eqnarray}%
Contracting even further, we are left with 
\begin{eqnarray}
\varphi _{abc}\varphi ^{abc} &=&42  \label{phiphi3} \\
\varphi _{abc}\psi _{m}^{\ \ abc} &=&0  \label{phipsi3} \\
\psi _{abcd}\psi _{m}^{\ \ \ bcd} &=&24g_{am}  \label{psipsi3} \\
\psi _{abcd}\psi ^{abcd} &=&168.  \label{psipsi4}
\end{eqnarray}%
The relations (\ref{phiphi3}) and (\ref{psipsi4}) both yield $\left\vert
\varphi \right\vert ^{2}=7$ in the inner product (\ref{forminp}). So in fact
we have 
\begin{equation}
V=\frac{1}{7}\int \varphi \wedge \ast \varphi  \label{phiwpsi}
\end{equation}%
where $V$ is the volume of the manifold $X$.

Now look in more detail at the decompositions (\ref{formdecompose}). We are
in particular interested in decompositions of $2$-forms and $3$-forms since
the decompositions for $4$-forms and $5$-forms are derived from these via
Hodge duality. From \cite{bryant-2003} and \cite{karigiannis-2005-57}, we
have 
\begin{eqnarray}
\Lambda _{7}^{2} &=&\left\{ \omega \lrcorner \varphi :\omega \ \text{a
vector field}\right\}  \label{om27} \\
\Lambda _{14}^{2} &=&\left\{ \alpha =\frac{1}{2}\alpha _{ab}dx^{a}\wedge
dx^{b}:\left( \alpha _{ab}\right) \in \mathfrak{g}_{2}\right\}  \label{om214}
\\
\Lambda _{1}^{3} &=&\left\{ f\varphi :f\ \text{a smooth function}\right\}
\label{om31} \\
\Lambda _{7}^{3} &=&\left\{ \omega \lrcorner \ast \varphi :\omega \ \text{a
vector field}\right\}  \label{om37} \\
\Lambda _{27}^{3} &=&\left\{ \chi \in \Omega ^{3}:\chi \wedge \varphi =0\ 
\text{and }\chi \wedge \ast \varphi =0\right\}  \label{om327}
\end{eqnarray}%
Following \cite{bryant-2003}, it is enough to consider what happens in $%
\mathbb{R}^{7}$ in order to understand these decompositions. Consider first
the Lie algebra $\mathfrak{so}\left( 7\right) $, which is the space of
antisymmetric $7\times 7$ matrices. For a vector $\omega \in \mathbb{R}^{7}$%
, define the map $\rho _{\varphi }:\mathbb{R}^{7}\longrightarrow \mathfrak{so%
}\left( 7\right) $ by $\rho _{\varphi }\left( \omega \right) =\omega
\lrcorner \varphi $, and this map is clearly injective. Conversely, define
the map $\tau _{\varphi }:$ $\mathfrak{so}\left( 7\right) \longrightarrow 
\mathbb{R}^{7}$ such that $\tau _{\varphi }\left( \alpha _{ab}\right) ^{c}=%
\frac{1}{6}\varphi _{\ \ ab}^{c}\alpha ^{ab}$. From (\ref{phiphi2}), we get
that 
\begin{equation*}
\tau _{\varphi }\left( \rho _{\varphi }\left( \omega \right) \right) =\omega
,
\end{equation*}%
so that $\tau _{\varphi }$ is a partial inverse of $\rho _{\varphi }$. Now
the Lie algebra $\mathfrak{g}_{2}$ can be defined as the kernel of $\tau
_{\varphi }$ \cite{karigiannis-2007}, that is%
\begin{equation}
\mathfrak{g}_{2}=\ker \tau _{\varphi }=\left\{ \alpha \in \mathfrak{so}%
\left( 7\right) :\varphi _{abc}\alpha ^{bc}=0\right\} .  \label{liealgg2}
\end{equation}%
This further implies that we get the following decomposition of $\mathfrak{so%
}\left( 7\right) $:%
\begin{equation}
\mathfrak{so}\left( 7\right) =\mathfrak{g}_{2}\oplus \rho _{\varphi }\left( 
\mathbb{R}^{7}\right) .  \label{so7decom}
\end{equation}%
The group $G_{2}$ acts via the adjoint representation on the $14$%
-dimensional vector space $\mathfrak{g}_{2}$ and via the natural, vector
representation on the $7$-dimensional space $\rho _{\varphi }\left( \mathbb{R%
}^{7}\right) $. This is a $G_{2}\,$-invariant irreducible decomposition of $%
\mathfrak{so}\left( 7\right) $ into the representations $\mathbf{7}$ and $%
\mathbf{14}$. Hence follows the decomposition of $\Lambda ^{2}$ (\ref%
{l1decom} and also the characterizations (\ref{om27}) and (\ref{om214}).

Following \cite{bryant-2003} again, let us look at $\Lambda _{27}^{3}$ in
more detail. Consider $Sym^{2}\left( \left( \mathbb{R}^{7}\right) ^{\ast
}\right) $ - the space of symmetric $2$-tensors and define a map $\mathrm{i}%
_{\varphi }:Sym^{2}\left( \left( \mathbb{R}^{7}\right) ^{\ast }\right)
\longrightarrow \Lambda ^{3}\left( \left( \mathbb{R}^{7}\right) ^{\ast
}\right) $ by 
\begin{equation}
\mathrm{i}_{\varphi }\left( h\right) _{abc}=h_{[a}^{d}\varphi _{bc]d}
\label{iphidef}
\end{equation}%
Clearly, 
\begin{equation*}
\mathrm{i}_{\varphi }\left( g\right) _{abc}=\varphi _{abc}.
\end{equation*}%
Now, we can decompose $Sym^{2}\left( \left( \mathbb{R}^{7}\right) ^{\ast
}\right) =\mathbb{R}g\oplus Sym_{0}^{2}\left( \left( \mathbb{R}^{7}\right)
^{\ast }\right) $ where $\mathbb{R}g$ is the set of symmetric tensors
proportional to the metric $g$ and $Sym_{0}^{2}\left( \left( \mathbb{R}%
^{7}\right) ^{\ast }\right) $ is the set of traceless symmetric tensors.
This is a $G_{2}$-invariant irreducible decomposition of $Sym^{2}\left(
\left( \mathbb{R}^{7}\right) ^{\ast }\right) $ into $1$-dimensional and $27$%
-dimensional components. The map $\mathrm{i}_{\varphi }$ is also $G_{2}$%
-invariant and is injective on each summand of this decomposition. Looking
at the first summand, we get that $\mathrm{i}_{\varphi }\left( \mathbb{R}%
g\right) =\Lambda _{1}^{3}\left( \left( \mathbb{R}^{7}\right) ^{\ast
}\right) $. Now look at the second summand and consider $\mathrm{i}_{\varphi
}\left( Sym_{0}^{2}\left( \left( \mathbb{R}^{7}\right) ^{\ast }\right)
\right) $. This is $27$-dimensional and irreducible, so by dimension count
it follows easily that $\mathrm{i}_{\varphi }\left( Sym_{0}^{2}\left( \left( 
\mathbb{R}^{7}\right) ^{\ast }\right) \right) =\Lambda _{27}^{3}\left(
\left( \mathbb{R}^{7}\right) ^{\ast }\right) $. All of this carries over to $%
3$-forms on our $G_{2}$ manifold $X$, and so we get 
\begin{equation}
\Lambda _{27}^{3}=\left\{ \chi \in \Lambda ^{3}:\chi
_{abc}=h_{[a}^{d}\varphi _{bc]d}\text{ for }h_{ab}~\text{traceless and
symmetric}\right\} .  \label{lamb327}
\end{equation}%
From the identities for contraction of $\varphi $ and $\ast \varphi $, it is
possible to see that this is equivalent to the description (\ref{om327}) of $%
\Lambda _{27}^{3}$. Thus we see that $1$-dimensional components correspond
to scalars, $7$-dimensional components correspond to vectors and $27$%
-dimensional components correspond to traceless symmetric matrices.

Now suppose we have $\chi \in \Lambda ^{3}$, then it is always useful to be
able to compute the different projections of $\chi $ into $\Lambda _{1}^{3}$%
, $\Lambda _{7}^{3}$ and $\Lambda _{27}^{3}$. Denote these projections by $%
\pi _{1}$, $\pi _{7}$ and $\pi _{27}$, respectively. As shown in Appendix 1,
we have the following relations%
\begin{eqnarray}
\pi _{1}\left( \chi \right) &=&a\varphi \ \text{where }a=\frac{1}{42}\left(
\chi _{abc}\varphi ^{abc}\right) =\frac{1}{7}\,\left\langle \chi ,\varphi
\right\rangle \ \text{and }\left\vert \pi _{1}\left( \chi \right)
\right\vert ^{2}=7a^{2}  \label{p1chi} \\
\pi _{7}\left( \chi \right) &=&\omega \lrcorner \ast \varphi \ \text{where }%
\omega ^{a}=-\frac{1}{24}\chi _{mnp}\psi ^{mnpa}\ \ \text{and }\left\vert
\pi _{7}\left( \chi \right) \right\vert ^{2}=4\left\vert \omega \right\vert
^{2}  \label{p7chi} \\
\pi _{27}\left( \chi \right) &=&\mathrm{i}_{\varphi }\left( h\right) \ \text{%
where }h_{ab}=\frac{3}{4}\chi _{mn\{a}\varphi _{b\}}^{\ \ mn}\ \text{and }%
\left\vert \pi _{27}\left( \chi \right) \right\vert ^{2}=\frac{2}{9}%
\left\vert h\right\vert ^{2}.  \label{p27chi}
\end{eqnarray}%
Here $\{a$ $b\}$ denotes the traceless symmetric part.

\section{$G_{2}$ manifolds in $M$-theory \label{mtheorysec}}

Special holonomy manifolds play a very important role in string and $M$%
-theory because of their relation to supersymmetry. In general, if we
compactify string or $M$-theory on a manifold of special holonomy $X$ the
preservation of supersymmetry is related to existence of covariantly
constant spinors (also known as parallel spinors). In fact, if all bosonic
fields except the metric are set to zero, and a supersymmetric vacuum
solution is sought, then in both string and $M$-theory, this gives precisely
the equation 
\begin{equation}
\nabla \xi =0  \label{covconstspinor}
\end{equation}%
for a spinor $\xi $. As lucidly explained in \cite{AcharyaGukov}, condition (%
\ref{covconstspinor}) on a spinor immediately implies special holonomy. Here 
$\xi $ is invariant under parallel transport, and is hence invariant under
the action of the holonomy group $Hol\left( X,g\right) $. This shows that
the spinor representation of $Hol\left( X,g\right) $ must contain the
trivial representation. For $Hol\left( X,g\right) =SO\left( n\right) $, this
is not possible since the spinor representation is reducible, so $Hol\left(
X,g\right) \subset SO\left( n\right) $. In particular, Calabi-Yau 3-folds
with $SU\left( 3\right) $ holonomy admit two covariantly constant spinors
and $G_{2}$ holonomy manifolds admit only one covariantly constant spinor.

Consider the bosonic action of eleven-dimensional supergravity \cite%
{Cremmer:1978km}, which is supposed to describe low-energy $M$-theory :

\begin{equation}
S=\frac{1}{2}\int d^{11}x\left( -\hat{g}\right) ^{\frac{1}{2}}R^{\left(
11\right) }-\frac{1}{4}\int G\wedge \ast G-\frac{1}{12}\int C\wedge G\wedge G
\label{sugraction}
\end{equation}%
where $\hat{g}$ is the metric on the $11$-dimensional space $M$ and $C$ is a 
$3$-form potential with field strength $G=dC$. From (\ref{sugraction}), the
equation of motion for $C$ is found to be 
\begin{equation}
d\ast G=\frac{1}{2}G\wedge G.  \label{dgeom}
\end{equation}%
Suppose we fix $M=M_{4}\times X$ where $M_{4}$ is the $4$-dimensional
Minkowski space and $X$ is a space with holonomy equal to $G_{2}$. Then $M$
is Ricci flat, so from Einstein's equation, $G$ has to vanish. However, it
turns out that the assumption that $G_{X}=\left. G\right\vert _{X}=0$ is not
an obvious one to make. In fact, as explained in \cite{Witten:1996md}, Dirac
quantization on $X$ gives a shifted quantization condition and gives the
statement 
\begin{equation}
\left[ \frac{G_{X}}{2\pi }\right] -\frac{\lambda }{2}\in H^{4}\left( X,%
\mathbb{Z}\right)  \label{quantcond}
\end{equation}%
where $\left[ \frac{G_{X}}{2\pi }\right] $ is the cohomology class of $\frac{%
G_{X}}{2\pi }$ and $\lambda =\frac{1}{2}$ $p_{1}\left( X\right) $ where $%
p_{1}\left( X\right) $ is the first Pontryagin class on $X$. So if $\lambda $
were not even in $H^{4}\left( X,\mathbb{Z}\right) $, then the ansatz $%
G_{X}=0 $ would not be consistent. Nonetheless, it was shown in \cite%
{Harvey:1999as} that if $X$ is a seven dimensional spin manifold (or in
particular $G_{2}$ holonomy manifold), then in fact $\lambda $ is even, and
setting $G_{X}=0$ is consistent.

So overall the simplest, Ricci-flat vacuum solutions are given by%
\begin{eqnarray}
\left\langle \hat{g}\right\rangle &=&\eta \times g_{7}  \label{vacghat} \\
\left\langle C\right\rangle &=&0  \label{vacc} \\
\left\langle G\right\rangle &=&0  \label{vacg}
\end{eqnarray}%
where $\left\langle \cdot \right\rangle $ denotes the vacuum expectation
value and $g_{7}$ is some metric with $G_{2}$ holonomy while $\eta $ is the
standard metric on the four dimensional Minkowski space. However, we know
that a $G_{2}$ structure and hence the metric $g_{7}$ is defined by a $G_{2}$%
-invariant $3$-form $\varphi _{0}$, so we have 
\begin{equation}
\left\langle \varphi \right\rangle =\varphi _{0}.  \label{vacphi}
\end{equation}%
Now consider small fluctuations about the vacuum, 
\begin{eqnarray}
\hat{g} &=&\left\langle \hat{g}\right\rangle +\delta \hat{g}
\label{vacgpert} \\
C &=&\left\langle C\right\rangle +\delta C=\delta C  \label{vaccpert} \\
\varphi &=&\left\langle \varphi \right\rangle +\delta \varphi =\varphi
_{0}+\delta \varphi  \label{vacphipert}
\end{eqnarray}

So a Kaluza-Klein ansatz for $C$ can be written as 
\begin{equation}
C=\sum_{N=1}^{b_{3}}c^{N}\left( x\right) \phi
_{N}+\sum_{I=1}^{b_{2}}A^{I}\left( x\right) \wedge \alpha _{I}
\label{Cansatz}
\end{equation}%
where $\left\{ \phi _{N}\right\} $ are a basis for harmonic $3$-forms on $X$%
, $\left\{ \alpha _{I}\right\} $ are a basis for harmonic $2$-forms on $X$, $%
c^{N}\left( x\right) $ are scalars on $M_{4}$ and $A^{I}\left( x\right) $
are $1$-forms on $M_{4}$ which describe the fluctuations of $C$. Also $b_{2}$
and $b_{3}$ are the Betti numbers of $X$. Since we assume that $X$ has
holonomy equal to $G_{2}$, $b_{1}=0$, so in (\ref{Cansatz}) we do not have a
contribution from harmonic $1$-forms on $X$. Now, deformations of the metric
on $X$ are encoded in the deformations of $\varphi $ and since $\varphi $ is
harmonic on $X$, we parametrize $\varphi $ as 
\begin{equation}
\varphi =\sum_{N=1}^{b_{3}}s^{N}\left( x\right) \phi _{N}.  \label{phiansatz}
\end{equation}%
Overall, in $4$ dimensions we get $b_{3}$ real scalars $c^{N}$ and $b_{3}$
real scalars $s^{N}$. Together these combine into $b_{3}$ massless complex
scalars $z^{N}$:%
\begin{equation}
z^{N}=\frac{1}{2}\left( s^{N}+ic^{N}\right) .  \label{complexz}
\end{equation}%
In the $4$-dimensional supergravity theory this gives $b_{3}$ massless
chiral superfields. The $1$-forms $A^{I}$ in (\ref{Cansatz}) give rise to $%
b_{2}$ massless Abelian gauge fields, and together with superpartners
arising from the gravitino fields, these form $b_{2}$ massless vector
superfields \cite{AcharyaGukov}. Thus overall, in four dimensions the
effective low energy theory is $\mathcal{N}=1$ supergravity coupled to $%
b_{2} $ abelian vector supermultiplets and $b_{3}$ massless chiral
supermultiplets. The physical theory is not very interesting from a
phenomenological point of view, since the gauge group is abelian and there
are no charged particles. However the combination (\ref{complexz}) proves to
be very useful for studying the moduli space of $G_{2}$ manifolds, since it
provides a natural, physically motivated complexification of the pure $G_{2}$
moduli space - something very similar to the complexified K\"{a}hler cone
used in the study of Calabi-Yau moduli spaces.

Let us now use our Kaluza-Klein ansatz to reduce the $11$-dimensional action
(\ref{sugraction}) to $4$ dimensions. Here we follow \cite{WittenBeasley},%
\cite{Gutowski:2001fm} and \cite{House:2004pm}. The term which interests us
is the kinetic term for the $z^{N}$. The kinetic term for the $c^{N}$, $%
L_{kin}\left( c\right) $ comes from the reduction of the $G\wedge \ast G$
term in (\ref{sugraction}). After switching to the Einstein frame by $g_{\mu
\nu }\longrightarrow V^{-1}g_{\mu \nu }$ we immediately see this gives us 
\begin{equation}
L_{kin}\left( c\right) =-\frac{1}{4V}\partial _{\mu }c^{M}\partial ^{\mu
}c^{N}\int_{X}\phi _{M}\wedge \ast \phi _{N}  \label{lkinc}
\end{equation}%
The kinetic term for the $s^{M}$ appears from the reduction of the $%
R^{\left( 11\right) }$ term in (\ref{sugraction}). This is less
straightforward that the derivation of $L_{kin}\left( c\right) $, but the
calculation was shown explicitly in \cite{House:2004pm}. From the general
properties of the Ricci scalar we can decompose the eleven-dimensional
Einstein-Hilbert action as 
\begin{equation}
\int d^{11}x\left( -\hat{g}\right) ^{\frac{1}{2}}R^{\left( 11\right) }=\int
d^{11}x\left( -\hat{g}\right) ^{\frac{1}{2}}V\left( R^{\left( 4\right)
}+R^{\left( 7\right) }+\frac{1}{4V}\left( \partial _{\mu }g_{mn}\partial
^{\mu }g^{mn}-\func{Tr}\left( \partial _{\mu }g\right) \func{Tr}\left(
\partial ^{\mu }g\right) \right) \right) .  \label{riccidecom}
\end{equation}%
Then, using deformation properties of the $G_{2}$ metric $g_{mn}$ from
section \ref{deformsec}, and switching to the Einstein frame $g_{\mu \nu
}\longrightarrow V^{-1}g_{\mu \nu }$, we eventually get 
\begin{equation}
L_{kin}\left( s\right) =-\frac{1}{4V}\partial _{\mu }s^{M}\partial ^{\mu
}s^{N}\int_{X}\phi _{M}\wedge \ast \phi _{N}.  \label{lkins}
\end{equation}%
The kinetic term of the dimensionally reduced action is in general given in
the Einstein frame by 
\begin{equation}
L_{kin}=-G_{M\bar{N}}\partial _{\mu }z^{M}\partial ^{\mu }\bar{z}^{N}.
\label{lkingen}
\end{equation}%
Comparing (\ref{lkingen}) with (\ref{lkinc}) and (\ref{lkins}), we can read
off the moduli space metric $G_{M\bar{N}}$ as 
\begin{equation}
G_{M\bar{N}}=\frac{1}{V}\int_{X}\phi _{M}\wedge \ast \phi _{\bar{N}}.
\label{modulimetric}
\end{equation}%
Note that the Hodge star implicitly depends on the coordinates $z^{M}$, so
this metric is quite non-trivial.

The bosonic part of fully reduced $4$-dimensional Lagrangian is given in
this case by \cite{Papadopoulos:1995da},\cite{Gutowski:2001fm} 
\begin{equation}
L=-G_{M\bar{N}}\partial _{\mu }z^{M}\partial ^{\mu }\bar{z}^{N}-\frac{1}{4}%
\func{Re}h_{IJ}F_{mn}^{I}F^{Jmn}+\frac{1}{4}\func{Im}h_{IJ}F_{mn}^{I}\ast
F^{Jmn}  \label{fulllag}
\end{equation}%
where $G_{M\bar{N}}$ is as in (\ref{modulimetric}), and 
\begin{equation*}
F_{mn}^{I}=\partial _{m}A_{n}^{I}-\partial _{n}A_{m}^{I}.
\end{equation*}%
The couplings $\func{Re}h_{IJ}$ and $\func{Im}h_{IJ}$ are given by 
\begin{eqnarray}
\func{Re}h_{IJ}\left( s\right) &=&\frac{1}{2}\int \alpha _{I}\wedge \ast
\alpha _{J}=-\frac{1}{2}s^{M}\int \alpha _{I}\wedge \alpha _{J}\wedge \phi
_{M}  \label{rehij} \\
\func{Im}h_{IJ}\left( c\right) &=&-\frac{1}{2}c^{M}\int \alpha _{I}\wedge
\alpha _{J}\wedge \phi _{M}  \label{imhij}
\end{eqnarray}%
To get the second equality in (\ref{rehij}) we have used that $%
H^{2}=H_{14}^{2}$ for manifolds with $G_{2}$ holonomy and that for a $2$%
-form $\alpha $, $2\ast \pi _{7}\left( \alpha \right) -\ast \pi _{14}\left(
\alpha \right) =\alpha \wedge \varphi $. Proof of this fact can be found in 
\cite{karigiannis-2005-57}.

\section{Deformations of $G_{2}$ structures \label{deformsec}}

\setcounter{equation}{0} As we already know, the $G_{2}$ structure on $X$
and the corresponding metric $g$ are all determined by the invariant $3$%
-form $\varphi $. Hence, deformations of $\varphi $ will induce deformations
of the metric. These deformations of metric will then also affect the
deformation of $\ast \varphi $. Since the relationship (\ref{metricdef})
between $g$ and $\varphi $ is non-linear, the resulting deformations of the
metric are highly non-trivial, and in general it is not possible to write
them down in closed form. However, as shown by Karigiannis in \cite%
{karigiannis-2005-57}, metric deformations can be made explicit when the $3$%
-form deformations are either in $\Lambda _{1}^{3}$ or $\Lambda _{7}^{3}$.
We now briefly review some of these results.

First suppose 
\begin{equation}
\tilde{\varphi}=f\varphi  \label{phitildeconf}
\end{equation}%
Then from (\ref{metriccomp1}) we get 
\begin{eqnarray}
\tilde{g}_{ab}\sqrt{\det \tilde{g}} &=&\frac{1}{144}\tilde{\varphi}_{amn}%
\tilde{\varphi}_{bpq}\tilde{\varphi}_{rst}\hat{\varepsilon}^{mnpqrst}  \notag
\\
&=&f^{3}g_{ab}\sqrt{\det g}  \label{p1g1}
\end{eqnarray}%
After taking the determinant on both sides, we obtain 
\begin{equation}
\det \tilde{g}=f^{\frac{14}{3}}\det g.  \label{p1detg1}
\end{equation}%
Substituting (\ref{p1detg1}) into (\ref{p1g1}), we finally get 
\begin{equation}
\tilde{g}_{ab}=f^{\frac{2}{3}}g_{ab}.  \label{p1g2}
\end{equation}%
and hence 
\begin{equation}
\tilde{\ast}\tilde{\varphi}=f^{\frac{4}{3}}\ast \varphi .  \label{p1sphi1}
\end{equation}%
Therefore, a scaling of $\varphi $ gives a conformal transformation of the
metric. hence deformations of $\varphi $ in the direction $\Lambda _{1}^{3}$
also give infinitesimal conformal transformation. Suppose $f=1+\varepsilon a$%
, then to fourth order in $\varepsilon $, we can write 
\begin{equation}
\tilde{\ast}\tilde{\varphi}=\left( \allowbreak 1+\frac{4}{3}a\varepsilon +%
\frac{2}{9}a^{2}\varepsilon ^{2}-\frac{4}{81}a^{3}\varepsilon ^{3}+\frac{5}{%
243}a^{4}\varepsilon ^{4}+O\left( \varepsilon ^{5}\right) \right) \ast
\varphi \allowbreak .  \label{p1sphi2}
\end{equation}

Now, suppose in general that $\tilde{\varphi}=\varphi +\varepsilon \chi $
for some $\chi \in \Lambda ^{3}$. Then using (\ref{metricdef}) for the
definition of the metric associated with $\tilde{\varphi}$, 
\begin{eqnarray}
\widetilde{\left\langle u,v\right\rangle }\widetilde{\mathrm{vol}} &=&\frac{1%
}{6}\left( u\lrcorner \tilde{\varphi}\right) \wedge \left( v\lrcorner \tilde{%
\varphi}\right) \wedge \tilde{\varphi}  \notag \\
&=&\frac{1}{6}\left( u\lrcorner \varphi \right) \wedge \left( v\lrcorner
\varphi \right) \wedge \varphi  \label{rhsdeform} \\
&&+\frac{1}{6}\varepsilon \left[ \left( u\lrcorner \chi \right) \wedge
\left( v\lrcorner \varphi \right) \wedge \varphi +\left( u\lrcorner \varphi
\right) \wedge \left( v\lrcorner \chi \right) \wedge \varphi +\left(
u\lrcorner \varphi \right) \wedge \left( v\lrcorner \varphi \right) \wedge
\chi \right]  \notag \\
&&+\frac{1}{6}\varepsilon ^{2}\left[ \left( u\lrcorner \chi \right) \wedge
\left( v\lrcorner \chi \right) \wedge \varphi +\left( u\lrcorner \varphi
\right) \wedge \left( v\lrcorner \chi \right) \wedge \chi +\left( u\lrcorner
\chi \right) \wedge \left( v\lrcorner \varphi \right) \wedge \chi \right] 
\notag \\
&&+\frac{1}{6}\varepsilon ^{3}\left( u\lrcorner \chi \right) \wedge \left(
v\lrcorner \chi \right) \wedge \chi  \notag
\end{eqnarray}%
After some manipulations, we can rewrite this as: 
\begin{eqnarray}
\widetilde{\left\langle u,v\right\rangle }\widetilde{\mathrm{vol}} &=&\frac{1%
}{6}\left( u\lrcorner \varphi \right) \wedge \left( v\lrcorner \varphi
\right) \wedge \varphi  \label{rhsdeform2} \\
&&+\frac{1}{2}\varepsilon \left[ \left( u\lrcorner \chi \right) \wedge \ast
\left( v\lrcorner \varphi \right) +\left( v\lrcorner \chi \right) \wedge
\ast \left( u\lrcorner \varphi \right) \right]  \notag \\
&&+\frac{1}{2}\varepsilon ^{2}\left( u\lrcorner \chi \right) \wedge \left(
v\lrcorner \chi \right) \wedge \varphi  \notag \\
&&+\frac{1}{6}\varepsilon ^{3}\left( u\lrcorner \chi \right) \wedge \left(
v\lrcorner \chi \right) \wedge \chi .  \notag
\end{eqnarray}%
Rewriting (\ref{rhsdeform2}) in local coordinates, we get%
\begin{equation}
\tilde{g}_{ab}\frac{\sqrt{\det \tilde{g}}}{\sqrt{\det g}}=g_{ab}+\frac{1}{2}%
\varepsilon \chi _{mn(a}\varphi _{b)}^{\ \ mn}+\frac{1}{8}\varepsilon
^{2}\chi _{amn}\chi _{bpq}\psi ^{mnpq}+\frac{1}{24}\varepsilon ^{3}\chi
_{amn}\chi _{bpq}\left( \ast \chi \right) ^{mnpq}  \label{rhsdeform3}
\end{equation}%
\qquad \qquad

Now suppose the deformation is in the $\Lambda _{7}^{3}$ direction. This
implies that 
\begin{equation}
\chi =\omega \lrcorner \ast \varphi  \label{p7chi1}
\end{equation}%
for some vector field $\omega $. Look at the first order term. From (\ref%
{p1chi}) and (\ref{p27chi}) we see that this is essentially a projection
onto $\Lambda _{1}^{3}\oplus \Lambda _{27}^{3}$ - the traceless part gives
the $\Lambda _{27}^{3}$ component and the trace gives the $\Lambda _{1}^{3}$
component. Hence this term vanishes for $\chi \in \Lambda _{7}^{3}$. For the
third order term, it is more convenient to study at it in (\ref{rhsdeform2}%
). By looking at 
\begin{equation*}
\omega \lrcorner \left( \left( u\lrcorner \omega \lrcorner \ast \varphi
\right) \wedge \left( v\lrcorner \omega \lrcorner \ast \varphi \right)
\wedge \ast \varphi \right) =0
\end{equation*}%
we immediately see that the third order term vanishes. So now we are left
with%
\begin{eqnarray}
\tilde{g}_{ab}\sqrt{\det \tilde{g}} &=&\left( g_{ab}+\frac{1}{8}\varepsilon
^{2}\omega ^{c}\omega ^{d}\psi _{camn}\psi _{dbpq}\psi ^{mnpq}\right) \sqrt{%
\det g}  \notag \\
&=&\left( g_{ab}\left( 1+\varepsilon ^{2}\left\vert \omega \right\vert
^{2}\right) -\varepsilon ^{2}\omega _{a}\omega _{b}\right) \sqrt{\det g}
\label{p7gabtil}
\end{eqnarray}%
where we have used the contraction identity for $\psi $ (\ref{psipsi2})
twice. Taking the determinant of (\ref{p7gabtil}) gives 
\begin{eqnarray}
\sqrt{\det \tilde{g}} &=&\left( 1+\varepsilon ^{2}\left\vert \omega
\right\vert ^{2}\right) ^{\frac{2}{3}}\sqrt{\det g}  \label{p7detg} \\
\tilde{g}_{ab} &=&\left( 1+\varepsilon ^{2}\left\vert \omega \right\vert
^{2}\right) ^{-\frac{2}{3}}\left( \left( g_{ab}\left( 1+\varepsilon
^{2}\left\vert \omega \right\vert ^{2}\right) -\varepsilon ^{2}\omega
_{a}\omega _{b}\right) \right)  \label{p7gab2}
\end{eqnarray}%
and eventually, 
\begin{equation}
\tilde{\ast}\tilde{\varphi}=\left( 1+\varepsilon ^{2}\left\vert \omega
\right\vert ^{2}\right) ^{-\frac{1}{3}}\left( \ast \varphi +\ast \varepsilon
\left( \omega \lrcorner \ast \varphi \right) +\varepsilon ^{2}\omega
\lrcorner \ast \left( \omega \lrcorner \varphi \right) \right) .
\label{p7starphi1}
\end{equation}%
The details of these last steps can be found in \cite{karigiannis-2005-57}.
Notice that to first order in $\varepsilon $, both $\sqrt{\det g}$ and $%
g_{ab}$ remain unchanged under this deformation. Now let us examine the last
term in (\ref{p7starphi1}) in more detail. Firstly, we have 
\begin{equation*}
\omega \lrcorner \ast \left( \omega \lrcorner \varphi \right) =\ast \left(
\omega ^{\flat }\wedge \left( \omega \lrcorner \varphi \right) \right)
\end{equation*}%
and 
\begin{eqnarray}
\left( \omega ^{\flat }\wedge \left( \omega \lrcorner \varphi \right)
\right) _{mnp} &=&3\omega _{\lbrack m}\omega ^{a}\varphi _{\left\vert
a\right\vert np]}  \notag \\
&=&3\mathrm{i}_{\varphi }\left( \omega \circ \omega \right)  \label{omomphi1}
\end{eqnarray}%
where $\left( \omega \circ \omega \right) _{ab}=\omega _{a}\omega _{b}$.
Therefore, in (\ref{p7starphi1}), this term gives $\Lambda _{1}^{4}$ and $%
\Lambda _{27}^{4}$ components. So, can write (\ref{p7starphi1}) as 
\begin{equation}
\tilde{\ast}\tilde{\varphi}=\left( 1+\varepsilon ^{2}\left\vert \omega
\right\vert ^{2}\right) ^{-\frac{1}{3}}\left( \left( 1+\frac{3}{7}%
\varepsilon ^{2}\left\vert \omega \right\vert ^{2}\right) \ast \varphi +\ast
\varepsilon \left( \omega \lrcorner \ast \varphi \right) +\varepsilon
^{2}\ast \mathrm{i}_{\varphi }\left( \left( \omega \circ \omega \right)
_{0}\right) \right) .  \label{p7starphi2}
\end{equation}%
Here $\left( \omega \circ \omega \right) _{0}$ denotes the traceless part of 
$\omega \circ \omega ,$ so that $\mathrm{i}_{\varphi }\left( \left( \omega
\circ \omega \right) _{0}\right) \in \Lambda _{27}^{3}$ and thus, in (\ref%
{p7starphi2}), the components in different representations are now
explicitly shown.

As we have seen above, in the cases when the deformations were in $\Lambda
_{1}^{3}$ or $\Lambda _{7}^{3}$ directions, there were some simplifications,
which make it possible to write down all results in a closed form. Now
however we will look at deformations in the $\Lambda _{27}^{3}$ directions,
and we will work to fourth order in $\varepsilon $. So suppose we have a
deformation 
\begin{equation*}
\tilde{\varphi}=\varphi +\varepsilon \chi
\end{equation*}%
where $\chi \in \Lambda _{27}^{3}$. Now let us set up some notation. Define 
\begin{eqnarray}
\tilde{s}_{ab} &=&\frac{1}{144}\frac{1}{\sqrt{\det g}}\tilde{\varphi}_{amn}%
\tilde{\varphi}_{bpq}\tilde{\varphi}_{rst}\hat{\varepsilon}^{mnpqrst}
\label{sabtilde} \\
&=&\tilde{g}_{ab}\frac{\sqrt{\det \tilde{g}}}{\sqrt{\det g}}
\label{sabtilde2}
\end{eqnarray}%
From (\ref{metriccomp1}), the untilded $s_{ab}$ is then just equal to $%
g_{ab} $. We can rewrite (\ref{sabtilde2}) as 
\begin{equation}
\left( g_{ab}+\delta g_{ab}\right) \frac{\sqrt{\det \tilde{g}}}{\sqrt{\det g}%
}=g_{ab}+\delta s_{ab}  \label{sabtilde3}
\end{equation}%
\qquad where $\delta g_{ab}$ is the deformation of the metric and $\delta
s_{ab}$ is the deformation of $s_{ab}$, which from (\ref{rhsdeform3}) is
given by%
\begin{equation}
\delta s_{ab}=\frac{1}{2}\varepsilon \chi _{mn(a}\varphi _{b)}^{\ \ mn}+%
\frac{1}{8}\varepsilon ^{2}\chi _{amn}\chi _{bpq}\psi ^{mnpq}+\frac{1}{24}%
\varepsilon ^{3}\chi _{amn}\chi _{bpq}\left( \ast \chi \right) ^{mnpq}.
\label{deltasab1}
\end{equation}%
Also introduce the following short-hand notation%
\begin{eqnarray}
s_{k} &=&\func{Tr}\left( \left( \delta s\right) ^{k}\right)  \label{skdef} \\
t_{k} &=&\func{Tr}\left( \left( \delta g\right) ^{k}\right)  \label{tkdef}
\end{eqnarray}%
where the trace is taken using the original metric $g$. From (\ref{deltasab1}%
), note that since $\chi \in \Lambda _{27}^{3}$, when taking the trace the
first order term vanishes, and hence $s_{1}$ is second-order in $\varepsilon 
$.

Further, after taking the trace of (\ref{sabtilde3}) using $g^{ab}$ and
rearranging, we have%
\begin{equation}
\sqrt{\frac{\det \tilde{g}}{\det g}}=\left( 1+\frac{1}{7}s_{1}\right) \left(
1+\frac{1}{7}t_{1}\right) ^{-1}  \label{detgdeform}
\end{equation}%
and hence 
\begin{equation}
\tilde{g}_{ab}=\tilde{s}_{ab}\left( 1+\frac{1}{7}t_{1}\right) \left( 1+\frac{%
1}{7}s_{1}\right) ^{-1}.  \label{metricdeform2}
\end{equation}%
As shown in Appendix B, we can also expand $\det \tilde{g}$ as 
\begin{eqnarray}
\frac{\det \tilde{g}}{\det g} &=&1+t_{1}+\frac{1}{2}\left(
t_{1}^{2}-t_{2}\right) +\frac{1}{6}\left( t_{1}^{3}-3t_{1}t_{2}+2t_{3}\right)
\label{detgtilde1} \\
&&+\frac{1}{24}\left(
t_{1}^{4}-6t_{1}^{2}t_{2}+3t_{2}^{2}+8t_{1}t_{3}-6t_{4}\right) +O\left(
\left\vert \delta g\right\vert ^{5}\right)  \notag
\end{eqnarray}%
and hence 
\begin{eqnarray}
\sqrt{\frac{\det \tilde{g}}{\det g}} &=&1+\frac{1}{2}t_{1}+\left( \frac{1}{8}%
t_{1}^{2}-\frac{1}{4}t_{2}\right) +\left( \frac{1}{48}t_{1}^{3}-\frac{1}{8}%
t_{1}t_{2}+\frac{1}{6}t_{3}\right)  \label{rdgtild1} \\
&&+\left( \frac{1}{384}t_{1}^{4}-\frac{1}{32}t_{1}^{2}t_{2}+\frac{1}{32}%
t_{2}^{2}\allowbreak +\frac{1}{12}t_{1}t_{3}-\frac{1}{8}t_{4}\right)
+O\left( \left\vert \delta g\right\vert ^{5}\right) .  \notag
\end{eqnarray}%
Thus we can equate (\ref{detgdeform}) and (\ref{rdgtild1}). Suppose $t_{1}$
is first order in $\varepsilon $. Then the only first order term in (\ref%
{rdgtild1}) is $\frac{1}{2}t_{1}$, but since $s_{1}$ is second-order, the
only first order term in (\ref{detgdeform}) is $-\frac{1}{7}t_{1}$. It
therefore follows that first order terms vanish, and so in fact $t_{1}$ is
also second-order in $\varepsilon $. This has profound consequences in that
we can ignore some of the terms in (\ref{rdgtild1}), as they give terms
higher than fourth order:

\begin{equation}
\sqrt{\frac{\det \tilde{g}}{\det g}}=1+\left( \frac{1}{2}t_{1}-\frac{1}{4}%
t_{2}\right) +\frac{1}{6}t_{3}+\left( \frac{1}{8}t_{1}^{2}-\frac{1}{8}%
t_{1}t_{2}+\frac{1}{32}t_{2}^{2}\allowbreak -\frac{1}{8}t_{4}\right)
+O\left( \varepsilon ^{5}\right) .  \label{rdgtild3}
\end{equation}

From (\ref{metricdeform2}) we can write down $\delta g_{ab}$ to fourth order
in $\varepsilon $ in terms of $t_{1}$ and quantities related to $\delta
s_{ab}$ and from this get $t_{2}$, $t_{3}$ and $t_{4}$ in terms of $t_{1}$
and $\delta s_{ab}$. So we have 
\begin{equation}
\delta g_{ab}=g_{ab}\left( \left( \frac{1}{7}t_{1}-\frac{1}{7}s_{1}\right)
+\left( \frac{1}{49}s_{1}^{2}-\frac{1}{49}s_{1}t_{1}\right) \right) +\delta
s_{ab}\left( \allowbreak 1+\left( \frac{1}{7}t_{1}-\frac{1}{7}s_{1}\right)
\right) +O\left( \varepsilon ^{5}\right)  \label{deltagab1}
\end{equation}%
and then from this, 
\begin{eqnarray}
t_{2} &=&s_{2}+\frac{1}{7}\left(
-s_{1}^{2}+t_{1}^{2}+2t_{1}s_{2}-2s_{1}s_{2}\right) +O\left( \varepsilon
^{5}\right)  \label{t2} \\
t_{3} &=&s_{3}+\frac{3}{7}\left( t_{1}s_{2}-s_{1}s_{2}\right) +O\left(
\varepsilon ^{5}\right)  \label{t3} \\
t_{4} &=&s_{4}+O\left( \varepsilon ^{5}\right) .  \label{t4}
\end{eqnarray}%
Substituting, (\ref{t2})-(\ref{t4}) into (\ref{rdgtild3}), we obtain%
\begin{equation}
\sqrt{\frac{\det \tilde{g}}{\det g}}=1+\left( -\frac{1}{4}s_{2}+\frac{1}{2}%
t_{1}\right) +\frac{1}{6}s_{3}+\left( -\frac{1}{8}s_{4}-\frac{1}{8}%
s_{2}t_{1}+\frac{1}{28}s_{1}^{2}+\frac{1}{32}s_{2}^{2}+\frac{5}{56}%
t_{1}^{2}\right) +O\left( \varepsilon ^{5}\right)  \label{rdgtild4}
\end{equation}%
After expanding (\ref{detgdeform}) to fourth order in $\varepsilon $ and
equating with (\ref{rdgtild4}), we are left with a quadratic equation for $%
t_{1}$:%
\begin{equation}
\frac{27}{392}t_{1}^{2}+t_{1}\left( \frac{9}{14}+\frac{1}{49}s_{1}-\frac{1}{8%
}s_{2}\right) +\left( -\frac{1}{7}s_{1}-\frac{1}{4}s_{2}+\frac{1}{6}s_{3}-%
\frac{1}{8}s_{4}+\frac{1}{28}s_{1}^{2}+\frac{1}{32}s_{2}^{2}\right) +O\left(
\varepsilon ^{5}\right) .  \label{t1quadeq}
\end{equation}%
Obviously there are two solutions, but turns out that one of them has a term
which is zero order in $\varepsilon $, so this does not fit our assumptions,
and hence we are only left with one solution, which to fourth order in $%
\varepsilon $ is given by%
\begin{equation}
t_{1}=\frac{2}{9}s_{1}+\frac{7}{18}s_{2}-\frac{7}{27}s_{3}+\left( \frac{7}{36%
}s_{4}+\frac{1}{81}s_{1}s_{2}-\frac{11}{162}s_{1}^{2}+\frac{7}{648}%
\allowbreak s_{2}^{2}\right) +O\left( \varepsilon ^{5}\right) .  \label{t1a}
\end{equation}%
Now that we have $t_{1}=\func{Tr}\left( \delta g\right) $, from (\ref%
{detgdeform}) we have 
\begin{equation}
\sqrt{\frac{\det \tilde{g}}{\det g}}=1+\left( \frac{1}{9}s_{1}-\frac{1}{18}%
s_{2}\right) +\frac{1}{27}s_{3}+\left( \frac{1}{162}s_{1}^{2}-\frac{1}{162}%
s_{1}s_{2}-\frac{1}{36}s_{4}+\frac{1}{648}s_{2}^{2}\right) +O\left(
\varepsilon ^{5}\right) .  \label{rdgtild5}
\end{equation}%
Using this and (\ref{sabtilde3}) we can immediately get the deformed metric.
The precise expression however is not very useful for us at this stage. What
we want is to be able to calculate the Hodge star with respect to the
deformed metric. So let $\alpha $ be a $3$-form, and consider the Hodge dual
of $\alpha $ with respect to the deformed metric:%
\begin{eqnarray*}
\left( \tilde{\ast}\alpha \right) _{mnpq} &=&\frac{1}{3!}\frac{1}{\sqrt{\det 
\tilde{g}}}\hat{\varepsilon}^{abcdrst}\tilde{g}_{ma}\tilde{g}_{nb}\tilde{g}%
_{pc}\tilde{g}_{qd}\alpha _{rst} \\
&=&\frac{\sqrt{\det g}}{\sqrt{\det \tilde{g}}}(\ast \alpha )^{abcd}\tilde{g}%
_{ma}\tilde{g}_{nb}\tilde{g}_{pc}\tilde{g}_{qd} \\
&=&\left( \frac{\det g}{\det \tilde{g}}\right) ^{\frac{5}{2}}(\ast \alpha
)^{abcd}\tilde{s}_{ma}\tilde{s}_{nb}\tilde{s}_{pc}\tilde{s}_{qd} \\
&=&\left( \frac{\det g}{\det \tilde{g}}\right) ^{\frac{5}{2}}\left( \left(
\ast \alpha \right) _{mnpq}+4\left( \ast \alpha \right) _{[mnp}^{\ \ \ \ \ \
d}\delta s_{q]d}+6\left( \ast \alpha \right) _{[mn}^{\ \ \ \ \ cd}\delta
s_{p\left\vert c\right\vert }\delta s_{q]d}\right. \\
&&\left. +4\left( \ast \alpha \right) _{[m}^{\ \ \ \ bcd}\delta
s_{n\left\vert b\right\vert }\delta s_{p\left\vert c\right\vert }\delta
s_{q]d}+\left( \ast \alpha \right) ^{abcd}\delta s_{am}\delta s_{bn}\delta
s_{cp}\delta s_{dq}\right)
\end{eqnarray*}%
From (\ref{rdgtild5}), the prefactor $\left( \frac{\det g}{\det \tilde{g}}%
\right) ^{\frac{5}{2}}$ is given to fourth order by 
\begin{equation}
\left( \frac{\det g}{\det \tilde{g}}\right) ^{\frac{5}{2}}=1+\left( -\frac{5%
}{9}s_{1}+\frac{5}{18}s_{2}\right) -\frac{5}{27}s_{3}+\left( \frac{5}{36}%
s_{4}-\frac{25}{162}s_{1}s_{2}+\frac{25}{162}s_{1}^{2}+\frac{25}{648}%
s_{2}^{2}\right) +O\left( \varepsilon ^{5}\right) .  \label{rdgtild5m5}
\end{equation}%
Finally, consider how $\ast \varphi $ deforms:%
\begin{eqnarray}
\left( \tilde{\ast}\tilde{\varphi}\right) _{mnpq} &=&\tilde{\ast}\varphi
_{mnpq}+\varepsilon \tilde{\ast}\chi _{mnpq}  \notag \\
&=&\left( \frac{\det g}{\det \tilde{g}}\right) ^{\frac{5}{2}}\left( \left(
\ast \varphi \right) _{mnpq}+4\left( \ast \varphi \right) _{[mnp}^{\ \ \ \ \
\ d}\delta s_{q]d}+6\left( \ast \varphi \right) _{[mn}^{\ \ \ \ \ cd}\delta
s_{p\left\vert c\right\vert }\delta s_{q]d}\right.  \label{starphi1} \\
&&+4\left( \ast \varphi \right) _{[m}^{\ \ \ \ bcd}\delta s_{n\left\vert
b\right\vert }\delta s_{p\left\vert c\right\vert }\delta s_{q]d}+\left( \ast
\varphi \right) ^{abcd}\delta s_{am}\delta s_{bn}\delta s_{cp}\delta s_{dq} 
\notag \\
&&+\varepsilon \left( \ast \chi \right) _{mnpq}+4\varepsilon \left( \ast
\chi \right) _{[mnp}^{\ \ \ \ \ \ d}\delta s_{q]d}+6\varepsilon \left( \ast
\chi \right) _{[mn}^{\ \ \ \ \ cd}\delta s_{p\left\vert c\right\vert }\delta
s_{q]d}  \notag \\
&&\left. +4\varepsilon \left( \ast \chi \right) _{[m}^{\ \ \ \ bcd}\delta
s_{n\left\vert b\right\vert }\delta s_{p\left\vert c\right\vert }\delta
s_{q]d}+O\left( \varepsilon ^{5}\right) \right)  \notag
\end{eqnarray}%
We ignored the last term, because overall it is at least fifth order.

So far, the only property of $\Lambda _{27}^{3}$ that we have used is that
it is orthogonal to $\varphi $, thus in fact, up to this point everything
applies to $\Lambda _{7}^{3}$ as well. Now however, let $\chi $ be of the
form 
\begin{equation}
\chi _{abc}=h_{[a}^{d}\varphi _{bc]d}  \label{chi27def}
\end{equation}%
where $h_{ab}$ is traceless and symmetric, so that $\chi \in \Lambda
_{27}^{3}$. Let us first introduce some further notation. Let $%
h_{1},h_{2},h_{3},h_{4}$ be traceless, symmetric matrices, and introduce the
following shorthand notation%
\begin{eqnarray}
\left( \varphi h_{1}h_{2}\varphi \right) _{mn} &=&\varphi
_{abm}h_{1}^{ad}h_{2}^{be}\varphi _{den}  \label{bphiab} \\
\varphi h_{1}h_{2}h_{3}\varphi  &=&\varphi
_{abc}h_{1}^{ad}h_{2}^{be}h_{3}^{cf}\varphi _{def}  \label{bphi} \\
\left( \psi h_{1}h_{2}h_{3}\psi \right) _{mn} &=&\psi _{abcm}\psi
_{defn}h_{1}^{ad}h_{2}^{be}h_{3}^{cf}  \label{bpsiab} \\
\psi h_{1}h_{2}h_{3}h_{4}\psi  &=&\psi _{abcm}\psi
_{defn}h_{1}^{ad}h_{2}^{be}h_{3}^{cf}h_{4}^{mn}  \label{bpsi}
\end{eqnarray}%
It is clear that all of these quantities are symmetric in the $h_{i}$ and
moreover $\left( \varphi h_{1}h_{2}\varphi \right) _{mn}$ and $\left( \psi
h_{1}h_{2}h_{3}\psi \right) _{mn}$ are both symmetric in indices $m$ and $n$%
. Then, it can be shown that 
\begin{eqnarray*}
\chi _{(a\left\vert mn\right\vert }\varphi _{b)}^{\ \ mn} &=&\frac{4}{3}%
h_{ab} \\
\chi _{amn}\chi _{bpq}\ast \varphi ^{mnpq} &=&-\frac{4}{7}\left\vert \chi
\right\vert ^{2}g_{ab}+\frac{16}{9}\left( h^{2}\right) _{\{ab\}}-\frac{4}{9}%
\left( \varphi hh\varphi \right) _{\{ab\}} \\
\chi _{amn}\chi _{bpq}\ast \chi ^{mnpq} &=&\frac{32}{189}\func{Tr}\left(
h^{3}\right) g_{ab}-\frac{8}{9}\left( \varphi hh^{2}\varphi \right) _{\{ab\}}
\end{eqnarray*}%
where as before $\left\{ a\ b\right\} $ denotes the traceless symmetric
part. Using this and (\ref{deltasab1}), we can now express $\delta s_{ab}$
in terms of $h$:%
\begin{eqnarray}
\delta s_{ab} &=&\frac{2}{3}\varepsilon h_{ab}+g_{ab}\left( -\frac{1}{14}%
\varepsilon ^{2}\left\vert \chi \right\vert ^{2}+\frac{4}{567}\varepsilon
^{3}\func{Tr}\left( h^{3}\right) \right)   \label{deltasabfull} \\
&&+\varepsilon ^{2}\left( \frac{2}{9}\left( h^{2}\right) _{\{ab\}}-\frac{1}{%
18}\left( \varphi hh\varphi \right) _{\{ab\}}\right) -\frac{\varepsilon ^{3}%
}{27}\left( \varphi hh^{2}\varphi \right) _{\{ab\}}  \notag
\end{eqnarray}%
and hence 
\begin{eqnarray}
s_{1} &=&\func{Tr}\left( \delta s\right) =-\frac{1}{2}\varepsilon
^{2}\left\vert \chi \right\vert ^{2}+\frac{4}{81}\varepsilon ^{3}\func{Tr}%
\left( h^{3}\right)   \label{s1full} \\
s_{2} &=&\func{Tr}\left( \delta s^{2}\right) =2\varepsilon ^{2}\left\vert
\chi \right\vert ^{2}+\varepsilon ^{3}\left( \frac{8}{27}\func{Tr}\left(
h^{3}\right) -\frac{2}{27}\left( \varphi hhh\varphi \right) \right) 
\label{s2full} \\
&&+\varepsilon ^{4}\left( -\frac{1}{16}\left\vert \chi \right\vert ^{4}+%
\frac{7}{162}\func{Tr}\left( h^{4}\right) -\frac{2}{81}\left( \varphi
hhh^{2}\varphi \right) +\frac{1}{324}\left( \psi hhhh\psi \right) \right)  
\notag \\
s_{3} &=&\func{Tr}\left( \delta s^{3}\right) =\frac{8}{27}\varepsilon ^{3}%
\func{Tr}\left( h^{3}\right) +\varepsilon ^{4}\left( -\frac{3}{2}\left\vert
\chi \right\vert ^{4}+\frac{8}{27}\func{Tr}\left( h^{4}\right) -\frac{2}{27}%
\left( \varphi hhh^{2}\varphi \right) \right) .  \label{s3full} \\
s_{4} &=&\func{Tr}\left( \delta s^{4}\right) =\frac{16}{81}\varepsilon ^{4}%
\func{Tr}\left( h^{4}\right)   \label{s4full}
\end{eqnarray}%
To get the full expression for $\tilde{\ast}\tilde{\varphi},$ (\ref{s1full}%
)-(\ref{s4full}) have to be substituted into the expression for the
prefactor $\left( \frac{\det g}{\det \tilde{g}}\right) ^{\frac{5}{2}}$ (\ref%
{rdgtild5m5}), and then both (\ref{rdgtild5m5}) and (\ref{deltasabfull})
have to be substituted into the expression for $\tilde{\ast}\tilde{\varphi}$
(\ref{starphi1}). Obviously, the expressions involved quickly become
absolutely gargantuan. Thankfully, we were able to use Maple and the freely
available package \emph{"Riegeom"} \cite{Riegeom} to help with these
calculations. After all the substitutions, the resulting expression still
has dozens of terms which are not of much use. In order for the expression
for $\tilde{\ast}\tilde{\varphi}$ to be useful, the terms in it have to be
separated according to which representation of $G_{2}$ they belong to. Thus
the final step is to apply projections onto $\Lambda _{1}^{4}$, $\Lambda
_{7}^{4}$ and $\Lambda _{27}^{4}$ (\ref{p1chi})-(\ref{p27chi}). When
applying these projections, many of the terms have $\varphi $ and $\psi $
contracted in some way, so the contraction identities (\ref{phiphi1})-(\ref%
{psipsi1}) have to be used to simplify the expressions. The package \emph{%
"Riegeom"} lacks the ability to make such substitutions, so a few simple
custom Maple programs based on \emph{"Riegeom" }had to be written in order
to facilitate these calculations. Overall, the expansion of $\tilde{\ast}%
\tilde{\varphi}$ to third order is%
\begin{eqnarray}
\tilde{\ast}\tilde{\varphi} &=&\ast \varphi -\varepsilon \ast \chi
+\varepsilon ^{2}\left( \frac{1}{6}\ast \mathrm{i}_{\varphi }\left( \left(
\phi hh\phi \right) _{0}\right) -\frac{1}{42}\left\vert \chi \right\vert
^{2}\ast \varphi \right)   \label{starphi27} \\
&&-\varepsilon ^{3}\left( \frac{2}{1701}\left( \varphi hhh\varphi \right)
\ast \varphi +\frac{5}{24}\left\vert \chi \right\vert ^{2}\ast \chi -\frac{1%
}{18}\ast \mathrm{i}_{\varphi }\left( h_{0}^{3}\right) +\frac{1}{36}\ast 
\mathrm{i}_{\varphi }\left( \left( \psi hhh\psi \right) _{0}\right) +\frac{1%
}{324}u\lrcorner \ast \varphi \right)   \notag \\
&&+O\left( \varepsilon ^{4}\right)   \notag
\end{eqnarray}%
where $\left( \phi hh\phi \right) _{0}$, $h_{0}^{3}$ and $\left( \psi
hhh\psi \right) _{0}$ denote the traceless parts of $\left( \phi hh\phi
\right) _{ab}$, $\left( h^{3}\right) _{ab}$ and $\left( \psi hhh\psi \right)
_{ab},$ respectively, and 
\begin{equation}
u^{a}=\psi _{\ mnp}^{a}\varphi _{rst}h^{mr}h^{ns}h^{pt}  \label{p27ua}
\end{equation}

Although above we did all calculations to fourth order, we will really only
need the expansion of $\tilde{\ast}\tilde{\varphi}$ to third order. However
for possible future reference here is the $G_{2}$ singlet piece of the
fourth order 
\begin{equation}
\left. \pi _{1}\left( \tilde{\ast}\tilde{\varphi}\right) \right\vert
_{\varepsilon ^{4}}=\frac{5}{13608}\left( \psi hhhh\psi \right) +\frac{25}{%
2016}\left\vert \chi \right\vert ^{4}-\frac{5}{6804}\func{Tr}\left(
h^{4}\right) .  \label{star4phi27p1}
\end{equation}%
In fact, using the homogeneity property of $\varphi \wedge \ast \varphi \,$%
it is possible to relate $\Lambda _{27}^{4}$ terms with a higher order $%
\Lambda _{1}^{4}$ term, so calculating higher order terms is also a way to
make sure that all the coefficients are consistent.

Now that we have expansions of $\tilde{\ast}\tilde{\varphi}$ for $1$- and $%
27 $-dimensional deformations, it is not difficult to combine them together.
Suppose we want to combine conformal transformation and $27$-dimensional
deformations. As in the case with $7$-dimensional deformations consider%
\begin{equation*}
\tilde{\varphi}=\hat{\varphi}+\varepsilon \chi
\end{equation*}%
where $\hat{\varphi}=f\varphi $ and $\chi \in \Lambda _{27}^{3}$. Consider
only up to second order in (\ref{starphi27}), 
\begin{equation*}
\tilde{\ast}\tilde{\varphi}=\hat{\ast}\hat{\varphi}-\varepsilon \hat{\ast}%
\chi +\varepsilon ^{2}\left( -\frac{1}{42}\widehat{\left\vert \chi
\right\vert }^{2}\hat{\ast}\hat{\varphi}+\frac{1}{6}\hat{\ast}\mathrm{i}_{%
\hat{\varphi}}\left( \left( \hat{\varphi}\hat{h}\hat{h}\hat{\varphi}\right)
_{0}\right) \right) +O\left( \varepsilon ^{3}\right) .
\end{equation*}%
Note that since $h_{ab}=\frac{3}{4}\chi _{mn\{a}\varphi _{b\}}^{\ \ mn}$, 
\begin{eqnarray*}
\hat{h}_{ab} &=&\frac{3}{4}\chi _{mn\{a}\hat{\varphi}_{b\}}^{\ \ mn}=\frac{3%
}{4}\hat{g}^{mr}\hat{g}^{ms}\chi _{mn\{a}\hat{\varphi}_{b\}rs}^{\ \ } \\
&=&f^{-\frac{1}{3}}h_{ab}
\end{eqnarray*}%
and hence 
\begin{eqnarray*}
\left( \hat{\varphi}\hat{h}\hat{h}\hat{\varphi}\right) _{ab} &=&\hat{\varphi}%
_{abm}\hat{\varphi}_{den}\hat{h}^{ad}\hat{h}^{be} \\
&=&f^{-\frac{4}{3}}\left( \varphi hh\varphi \right) _{ab}.
\end{eqnarray*}%
Moreover, 
\begin{equation*}
\mathrm{i}_{\hat{\varphi}}\left( \left( \hat{\varphi}\hat{h}\hat{h}\hat{%
\varphi}\right) _{0}\right) =f^{-1}\mathrm{i}_{\varphi }\left( \left(
\varphi hh\varphi \right) _{0}\right) .
\end{equation*}%
Therefore, overall, 
\begin{equation}
\tilde{\ast}\tilde{\varphi}=f^{\frac{4}{3}}\ast \varphi -\varepsilon f^{%
\frac{1}{3}}\ast \chi +\varepsilon ^{2}\left( -\frac{1}{42}f^{-\frac{2}{3}%
}\left\vert \chi \right\vert ^{2}\ast \varphi +\frac{1}{6}f^{-\frac{2}{3}%
}\ast \mathrm{i}_{\varphi }\left( \left( \varphi hh\varphi \right)
_{0}\right) \right) +O\left( \varepsilon ^{3}\right) .  \label{sphiconfp27}
\end{equation}%
Let $f=1+\varepsilon a$, and expand in powers of $\varepsilon $ to third
order to get 
\begin{eqnarray}
\tilde{\ast}\tilde{\varphi} &=&\ast \varphi +\varepsilon \left( \frac{4}{3}%
a\ast \varphi -\ast \chi \right) +\varepsilon ^{2}\left( \left( \frac{2}{9}%
a^{2}-\frac{1}{42}\left\vert \chi \right\vert ^{2}\right) \ast \varphi -%
\frac{1}{3}a\ast \chi +\frac{1}{6}\ast \mathrm{i}_{\varphi }\left( \left(
\varphi hh\varphi \right) _{0}\right) \right)  \label{sphip1p27} \\
&&+\varepsilon ^{3}\left( \left( \frac{1}{63}a\left\vert \chi \right\vert
^{2}-\frac{4}{81}a^{3}\right) \ast \varphi -\frac{1}{9}a\ast \mathrm{i}%
_{\varphi }\left( \left( \varphi hh\varphi \right) _{0}\right) +\left( \frac{%
1}{9}a^{2}-\frac{5}{24}\left\vert \chi \right\vert ^{2}\right) \ast \chi
\right)  \notag \\
&&+\varepsilon ^{3}\left( \frac{1}{18}\ast \mathrm{i}_{\varphi }\left(
h_{0}^{3}\right) -\frac{1}{36}\ast \mathrm{i}_{\varphi }\left( \left( \psi
hhh\psi \right) _{0}\right) -\frac{2}{1701}\left( \varphi hhh\varphi \right)
\ast \varphi -\frac{1}{324}u\lrcorner \ast \varphi \right) +O\left(
\varepsilon ^{4}\right)  \notag
\end{eqnarray}

\section{Moduli space}

\setcounter{equation}{0}In section \ref{mtheorysec} we described how $M$%
-theory can be used to give a natural complexification of the $G_{2}$ moduli
space - denote this space by $\mathcal{M}_{\mathbb{C}}$. The metric (\ref%
{modulimetric}) on $\mathcal{M}_{\mathbb{C}}$ arises naturally from the
Kaluza-Klein reduction of the $M$-theory action. As shown in \cite%
{WittenBeasley}, it turns out that this metric is in fact K\"{a}hler, with
the K\"{a}hler potential $K$ given by 
\begin{equation}
K=-3\log V.  \label{kahlerpot}
\end{equation}%
where as before, $V$ is the volume of $X$ 
\begin{equation*}
V=\frac{1}{7}\int \varphi \wedge \ast \varphi .
\end{equation*}%
Note that in sometimes $K$ is given with a different normalization factor.
Here we follow \cite{WittenBeasley}, but in \cite{Gutowski:2001fm} and \cite%
{karigiannis-2007a}, in particular, a different convention is used.

Let us show that $K$ is indeed the K\"{a}hler potential for $G_{M\bar{N}}$.
Clearly, $V,$ $K$ and $G_{M\bar{N}}$ only depend on the parameters $s^{M}$
for the $G_{2}$ $3$-form - that is, only the real part $s^{M}$ of the
complex coordinates $z^{M}$ on $\mathcal{M}_{\mathbb{C}}$. So let us for now
just look at the $s^{M}$ derivatives. Note that under a scaling $%
s^{M}\longrightarrow \lambda s^{M}$, $\varphi $ scales as $\varphi
\longrightarrow \lambda \varphi $ and from (\ref{p1sphi1}), $\ast \varphi $
scales as $\ast \varphi \longrightarrow \lambda ^{\frac{4}{3}}\ast \varphi $%
, and so $V$ scales as 
\begin{equation*}
V\longrightarrow \lambda ^{\frac{7}{3}}V.
\end{equation*}%
So $V$ is homogeneous of order $\frac{7}{3}$ in the $s^{M}$, and hence%
\begin{eqnarray*}
s^{M}\frac{\partial V}{\partial s^{M}} &=&\frac{7}{3}V \\
&=&\frac{1}{3}\int s^{M}\phi _{M}\wedge \ast \varphi 
\end{eqnarray*}%
and thus, 
\begin{equation}
\frac{\partial V}{\partial s^{M}}=\frac{1}{3}\int \phi _{M}\wedge \ast
\varphi .  \label{dvdsm}
\end{equation}%
Hence, 
\begin{equation}
\frac{\partial K}{\partial s^{M}}=-\frac{1}{V}\int \phi _{M}\wedge \ast
\varphi .  \label{dkdsm}
\end{equation}%
Here the dependence on the $s^{M}$ is encoded in $V$ and in $\ast \varphi $,
which depends non-linearly on the $s^{M}$. Thus we have, 
\begin{eqnarray*}
\frac{\partial ^{2}K}{\partial z^{M}\partial \bar{z}^{N}} &=&\frac{3}{V^{2}}%
\frac{\partial V}{\partial s^{M}}\frac{\partial V}{\partial s^{N}}-\frac{3}{V%
}\frac{\partial ^{2}V}{\partial s^{M}\partial s^{N}} \\
&=&\frac{1}{3}\frac{1}{V^{2}}\left( \int \phi _{M}\wedge \ast \varphi
\right) \left( \int \phi _{N}\wedge \ast \varphi \right) -\frac{1}{V}\int
\phi _{(M}\wedge \partial _{N)}\left( \ast \varphi \right) .
\end{eqnarray*}%
As we know from section \ref{deformsec} , the first derivative of $\ast
\varphi $ is given by 
\begin{equation}
\partial _{N}\left( \ast \varphi \right) =\frac{4}{3}\ast \pi _{1}\left(
\phi _{N}\right) +\ast \pi _{7}\left( \phi _{N}\right) -\ast \pi _{27}\left(
\phi _{N}\right) .  \label{sphi1der}
\end{equation}%
so therefore, 
\begin{eqnarray*}
\int \phi _{(M}\wedge \partial _{N)}\left( \ast \varphi \right)  &=&\frac{4}{%
3}\int \left( \pi _{1}\left( \varphi _{M}\right) \wedge \ast \pi _{1}\left(
\varphi _{N}\right) \right) +\int \left( \pi _{7}\left( \varphi _{M}\right)
\wedge \ast \pi _{7}\left( \varphi _{N}\right) \right)  \\
&&-\int \left( \pi _{27}\left( \varphi _{M}\right) \wedge \ast \pi
_{27}\left( \varphi _{N}\right) \right) 
\end{eqnarray*}%
Also using (\ref{p1chi}), we get 
\begin{equation}
\frac{1}{3}\frac{1}{V^{2}}\left( \int \phi _{M}\wedge \ast \varphi \right)
\left( \int \phi _{N}\wedge \ast \varphi \right) =\frac{7}{3}\frac{1}{V}\int
\pi _{1}\left( \varphi _{M}\right) \wedge \ast \pi _{1}\left( \varphi
_{N}\right)   \label{p1p1a}
\end{equation}%
Thus overall, 
\begin{equation}
\frac{\partial ^{2}K}{\partial z^{M}\partial \bar{z}^{N}}=\frac{1}{V}\left(
\int \left( \pi _{1}\left( \varphi _{M}\right) \wedge \ast \pi _{1}\left(
\varphi _{N}\right) \right) -\int \left( \pi _{7}\left( \varphi _{M}\right)
\wedge \ast \pi _{7}\left( \varphi _{N}\right) \right) +\int \left( \pi
_{27}\left( \varphi _{M}\right) \wedge \ast \pi _{27}\left( \varphi
_{N}\right) \right) \right) .  \label{d2kss}
\end{equation}%
Note that if $Hol\left( X\right) =G_{2}$ then all the seven-dimensional
components vanish, and hence we get 
\begin{equation}
\frac{\partial ^{2}K}{\partial z^{M}\partial \bar{z}^{N}}=\frac{1}{V}%
\int_{X}\phi _{M}\wedge \ast \phi _{\bar{N}}=G_{M\bar{N}}\text{,}
\label{d2kzz}
\end{equation}%
as claimed. Since the negative definite part of (\ref{d2kss}) vanishes, the
resulting metric is positive definite.

In general, there is at least one other good candidate for the metric on the 
$G_{2}$ moduli space. The Hessian of $V$, rather than of $\log V$, can be
used as a K\"{a}hler potential and gives a metric with signature $\left(
1,b_{27}^{3}\right) $. This metric is in particular used in \cite%
{Hitchin:2000jd} and \cite{karigiannis-2007a}. There are some advantages to
using $V$ as the K\"{a}hler potential, because some computations give more
elegant results. However if we use the supergravity action as a starting
point for the study of the moduli space, our choice of the K\"{a}hler
potential is very natural.

Now we have a complex manifold $\mathcal{M}_{\mathbb{C}}$, equipped with the
K\"{a}hler metric $G_{M\bar{N}}$, so it is now interesting to study the
properties of this metric, and the geometry which it gives. We will use the
metric $G_{M\bar{N}}$ to calculate the associated curvature tensor $\mathcal{%
R}_{M\bar{N}P\bar{Q}}$ of the manifold $\mathcal{M}_{\mathbb{C}}.$ Note that
calculation of the curvature of the moduli space but for a different choice
of metric is done in \cite{KarigiannisLin}.

Let us introduce local special coordinates on $\mathcal{M}_{\mathbb{C}}$.
Let $\phi _{0}=a\varphi $ and $\phi _{\mu }\in \Lambda _{27}^{3}$ for $\mu
=1,...,b_{27}^{3}$, so $s^{0}$ defines directions parallel to $\varphi $ and 
$s^{\mu }$ define directions in $\Lambda _{27}^{3}$. Since our metric is K%
\"{a}hler, the expression for $\mathcal{R}_{M\bar{N}P\bar{Q}}$ is given by 
\begin{equation}
\mathcal{R}_{\bar{K}L\bar{M}N}=\partial _{\bar{M}}\partial _{N}\partial
_{L}\partial _{\bar{K}}K-G^{R\bar{S}}\left( \partial _{\bar{M}}\partial
_{R}\partial _{\bar{K}}K\right) \left( \partial _{N}\partial _{L}\partial _{%
\bar{S}}K\right) .  \label{modcurv1}
\end{equation}%
Also define 
\begin{equation}
A_{MNR}=\frac{\partial ^{3}K}{\partial z^{M}\partial z^{N}\partial z^{R}}
\label{yuk1}
\end{equation}%
so that we can rewrite (\ref{modcurv1}) as 
\begin{equation}
\mathcal{R}_{\bar{K}L\bar{M}N}=\partial _{\bar{M}}\partial _{N}\partial
_{L}\partial _{\bar{K}}K-G^{R\bar{S}}A_{\bar{M}R\bar{K}}A_{NL\bar{S}}.
\label{modcurv2}
\end{equation}%
Now it only remains to work out the third and fourth derivatives of $K$.
Starting from (\ref{dkdsm}) we find that 
\begin{eqnarray}
A_{MNR} &=&-\frac{1}{V}\int \phi _{M}\wedge \frac{\partial ^{2}}{\partial
s^{N}\partial s^{R}}\left( \ast \varphi \right) +\frac{1}{V^{2}}\left( \int
\phi _{(M}\wedge \ast \varphi \right) \left( \int \phi _{N}\wedge \frac{%
\partial }{\partial s^{R)}}\left( \ast \varphi \right) \right) 
\label{amnk1} \\
&&-\frac{2}{9V^{3}}\left( \int \phi _{M}\wedge \ast \varphi \right) \left(
\int \phi _{N}\wedge \ast \varphi \right) \left( \int \phi _{R}\wedge \ast
\varphi \right)   \notag
\end{eqnarray}%
and from the power series expansion of $\ast \varphi $ (\ref{sphip1p27}), we
can extract the higher derivatives of $\ast \varphi $: 
\begin{subequations}
\label{sphiderivs}
\begin{eqnarray}
\partial _{0}\partial _{0}\left( \ast \varphi \right)  &=&\frac{4}{9}%
a^{2}\ast \varphi \ \ \ \ \ \ \ \partial _{0}\partial _{0}\partial
_{0}\left( \ast \varphi \right) =-\frac{8}{27}a^{3}\ast \varphi 
\label{sphider1} \\
\partial _{0}\partial _{\mu }\left( \ast \varphi \right)  &=&-\frac{1}{3}%
a\ast \phi _{\mu }\ \ \ \ \ \ \partial _{0}\partial _{0}\partial _{\mu
}\left( \ast \varphi \right) =\frac{2}{9}a^{2}\ast \phi _{\mu }\ \ \ 
\label{sphider2} \\
\partial _{\mu }\partial _{\nu }\left( \ast \varphi \right)  &=&-\frac{1}{21}%
\left\langle \phi _{\mu },\phi _{\nu }\right\rangle \ast \varphi +\frac{1}{3}%
\ast \mathrm{i}_{\varphi }\left( \left( \varphi h_{\mu }h_{\nu }\varphi
\right) _{0}\right)   \label{sphider3} \\
\partial _{0}\partial _{\mu }\partial _{\nu }\left( \ast \varphi \right)  &=&%
\frac{2}{63}a\left\langle \phi _{\mu },\phi _{\nu }\right\rangle \ast
\varphi -\frac{2}{9}a\ast \mathrm{i}_{\varphi }\left( \left( \varphi h_{\mu
}h_{\nu }\varphi \right) _{0}\right)   \label{sphider4} \\
\partial _{\mu }\partial _{\nu }\partial _{\kappa }\left( \ast \varphi
\right)  &=&-\frac{5}{4}\left\langle \phi _{\mu },\phi _{\nu }\right\rangle
\ast \phi _{\kappa }+\frac{1}{3}\ast \mathrm{i}_{\varphi }\left( (h_{\mu
}h_{\nu }h_{\kappa })_{0}\right) -\frac{1}{6}\ast \mathrm{i}_{\varphi
}\left( \left( \psi h_{\mu }h_{\nu }h_{\kappa }\psi \right) _{0}\right) 
\label{sphider5} \\
&&-\frac{4}{567}\left( \varphi h_{\mu }h_{\nu }h_{\kappa }\varphi \right)
\ast \varphi   \notag
\end{eqnarray}%
where $h_{\mu }$,$h_{\nu }$ and $h_{\kappa }$ are traceless symmetric
matrices corresponding to the $3$-forms $\phi _{\mu }$,$\varphi _{\nu }$ and 
$\phi _{\kappa }$, respectively. Using these expressions, we can now write
down all the components of $A_{MNR}$: 
\end{subequations}
\begin{subequations}
\label{Amnkp1p27}
\begin{eqnarray}
A_{\bar{0}0\bar{0}} &=&-14a^{3}  \label{a000} \\
A_{\bar{0}0\bar{\mu}} &=&0  \label{a00m} \\
A_{\bar{0}\mu \bar{\nu}} &=&-\frac{2a}{V}\int \phi _{\mu }\wedge \ast \phi _{%
\bar{\nu}}=-2aG_{\mu \bar{\nu}}  \label{a0mn} \\
A_{\bar{\mu}\nu \bar{\rho}} &=&-\frac{2}{27V}\int \left( \varphi h_{\bar{\mu}%
}h_{\nu }h_{\bar{\rho}}\varphi \right) dV  \label{amnr}
\end{eqnarray}%
Now also look at the fourth derivative of $K$. From (\ref{sphiderivs}), we
get 
\end{subequations}
\begin{subequations}
\label{k4thder}
\begin{eqnarray}
\frac{\partial ^{4}K}{\partial z^{0}\partial \bar{z}^{0}\partial
z^{0}\partial \bar{z}^{0}} &=&42a^{4}  \label{k0000} \\
\frac{\partial ^{4}K}{\partial z^{0}\partial \bar{z}^{0}\partial
z^{0}\partial \bar{z}^{\mu }} &=&0  \label{k000m} \\
\frac{\partial ^{4}K}{\partial z^{0}\partial \bar{z}^{0}\partial z^{\mu
}\partial \bar{z}^{\nu }} &=&\frac{4}{3}\frac{a^{2}}{V}\int \phi _{\mu
}\wedge \ast \phi _{\bar{\nu}}=\frac{4}{3}a^{2}G_{\mu \bar{\nu}}
\label{k00mn} \\
\frac{\partial ^{4}K}{\partial z^{0}\partial \bar{z}^{\mu }\partial z^{\nu
}\partial \bar{z}^{\rho }} &=&\frac{2}{9}\frac{a}{V}\int \left( \varphi
h_{\mu }h_{\nu }h_{\rho }\varphi \right) \mathrm{vol}=-3aA_{\bar{\mu}\nu 
\bar{\rho}}  \label{k0mnr} \\
\frac{\partial ^{4}K}{\partial z^{\kappa }\partial \bar{z}^{\mu }\partial
z^{\nu }\partial \bar{z}^{\rho }} &=&\frac{1}{3}\left( G_{\bar{\mu}\nu
}G_{\kappa \bar{\rho}}+G_{\bar{\mu}\kappa }G_{\nu \bar{\rho}}\right) +\frac{1%
}{3}\frac{1}{V^{2}}\int \phi _{\kappa }\wedge \ast \phi _{\nu }\int \phi _{%
\bar{\mu}}\wedge \ast \phi _{\bar{\rho}}  \label{kkmnr} \\
&&+\frac{1}{27V}\int \left( \left( \psi h_{\kappa }h_{\bar{\mu}}h_{\nu }h_{%
\bar{\rho}}\psi \right) -2\func{Tr}\left( h_{\kappa }h_{\bar{\mu}}h_{\nu }h_{%
\bar{\rho}}\right) +\frac{5}{3}\func{Tr}\left( h_{(\kappa }h_{\bar{\mu}%
}\right) \func{Tr}\left( h_{\nu }h_{\bar{\rho})}\right) \right) \mathrm{vol}
\notag
\end{eqnarray}%
Note that it can be shown using the identity (\ref{psipsi0}) that 
\end{subequations}
\begin{equation*}
\psi hhhh\psi =12\left( \varphi h^{2}hh\varphi \right) +3\func{Tr}\left(
h^{2}\right) ^{2}-6\func{Tr}\left( h^{4}\right) 
\end{equation*}%
\qquad 

Now define 
\begin{equation}
C_{MN}=\frac{\partial ^{2}K}{\partial z^{M}\partial z^{N}}  \label{cmn0}
\end{equation}%
This is the second derivative of $K$ but with pure indices, rather than the
derivative with mixed indices which gives the metric $G_{M\bar{N}}$. Note
that since $K=K\left( \func{Im}z\right) $, we have 
\begin{equation}
\frac{\partial ^{2}K}{\partial z^{M}\partial z^{N}}=\frac{\partial ^{2}K}{%
\partial z^{M}\partial \bar{z}^{N}}  \label{gmncmn}
\end{equation}%
so numerically, $C_{MN}$ and $G_{M\bar{N}}$ are in fact equal, and in
particular, 
\begin{equation}
C_{\mu \nu }=\frac{1}{V}\int \phi _{\mu }\wedge \ast \phi _{\nu }
\label{cmn}
\end{equation}%
So while $C_{MN}$ is not technically part of the metric, it inherits some
similar properties. This happens due to the fact that while the
complexification of the moduli space comes naturally, the holomorphic
structure is artificial to some extent, because the $G_{2}$ and $C$-field
moduli do not really mix. Using this definition, we can rewrite (\ref{kkmnr}%
) as 
\begin{eqnarray*}
\frac{\partial ^{4}K}{\partial z^{\kappa }\partial \bar{z}^{\mu }\partial
z^{\nu }\partial \bar{z}^{\rho }} &=&\frac{1}{3}\left( G_{\bar{\mu}\nu
}G_{\kappa \bar{\rho}}+G_{\bar{\mu}\kappa }G_{\nu \bar{\rho}}\right) +\frac{1%
}{3}C_{\bar{\mu}\bar{\rho}}C_{\kappa \nu } \\
&&-\frac{1}{27V}\int \left( 2\func{Tr}\left( h_{\kappa }h_{\bar{\mu}}h_{\nu
}h_{\bar{\rho}}\right) -\left( \psi h_{\kappa }h_{\bar{\mu}}h_{\nu }h_{\bar{%
\rho}}\psi \right) -\frac{5}{3}\func{Tr}\left( h_{(\kappa }h_{\bar{\mu}%
}\right) \func{Tr}\left( h_{\nu }h_{\bar{\rho})}\right) \right) \mathrm{vol}
\end{eqnarray*}%
Taking into account that $G^{0\bar{0}}=\frac{1}{7a^{2}}$ and $G^{0\bar{\mu}%
}=0$, we have enough information to be able to write down the full
expressions for the components of the curvature tensor: 
\begin{eqnarray}
\mathcal{R}_{0\bar{0}0\bar{0}} &=&14a^{4}  \label{R0000} \\
\mathcal{R}_{0\bar{0}0\bar{\mu}} &=&0  \label{R000m} \\
\mathcal{R}_{0\bar{0}\mu \bar{\nu}} &=&2a^{2}G_{\mu \bar{\nu}}  \label{R00mn}
\\
\mathcal{R}_{0\bar{\mu}\nu \bar{\rho}} &=&-A_{\bar{\mu}\nu \bar{\rho}}a
\label{R0mnr} \\
\mathcal{R}_{\kappa \bar{\mu}\nu \bar{\rho}} &=&\frac{1}{3}\left( G_{\bar{\mu%
}\nu }G_{\kappa \bar{\rho}}+G_{\bar{\mu}\kappa }G_{\nu \bar{\rho}}\right)
-G^{\tau \bar{\sigma}}A_{\bar{\mu}\tau \bar{\rho}}A_{\kappa \nu \bar{\sigma}%
}-\frac{5}{21}C_{\bar{\mu}\bar{\rho}}C_{\kappa \nu }  \label{Rkmnr} \\
&&+\frac{1}{27V}\int \left( \left( \psi h_{\kappa }h_{\bar{\mu}}h_{\nu }h_{%
\bar{\rho}}\psi \right) -2\func{Tr}\left( h_{\kappa }h_{\bar{\mu}}h_{\nu }h_{%
\bar{\rho}}\right) +\frac{5}{3}\func{Tr}\left( h_{(\kappa }h_{\bar{\mu}%
}\right) \func{Tr}\left( h_{\nu }h_{\bar{\rho})}\right) \right) \mathrm{vol}
\notag
\end{eqnarray}%
Let us look at more detail at the expression for $A_{\mu \bar{\nu}\bar{\rho}}
$: 
\begin{eqnarray*}
A_{\bar{\mu}\nu \bar{\rho}} &=&-\frac{2}{27V}\int \varphi h_{\bar{\mu}%
}h_{\nu }h_{\bar{\rho}}\varphi \mathrm{vol} \\
&=&-\frac{2}{27V}\int \varphi _{abc}\varphi _{mnp}h_{\bar{\mu}}^{am}h_{\nu
}^{bn}h_{\bar{\rho}}^{cp}\mathrm{vol}
\end{eqnarray*}%
Define $h_{\mu }^{a}=h_{\mu \ m}^{\ a}dx^{m}.$ Then 
\begin{equation*}
\varphi _{abc}\varphi _{mnp}h_{\bar{\mu}}^{am}h_{\nu }^{bn}h_{\bar{\rho}%
}^{cp}\mathrm{vol}=6\varphi _{abc}h_{\bar{\mu}}^{a}\wedge h_{\nu }^{b}\wedge
h_{\bar{\rho}}^{c}\wedge \ast \varphi 
\end{equation*}%
and so, 
\begin{equation}
A_{\bar{\mu}\nu \bar{\rho}}=-\frac{4}{9V}\int \varphi _{abc}h_{\bar{\mu}%
}^{a}\wedge h_{\nu }^{b}\wedge h_{\bar{\rho}}^{c}\wedge \ast \varphi .
\label{amnryuk}
\end{equation}%
This is the precise analogue of the Yukawa coupling which is defined on the
Calabi-Yau moduli space. Similar expressions have appeared previously in 
\cite{Lee:2002fa},\cite{deBoer:2005pt} and \cite{karigiannis-2007a}.
Similarly, we can write 
\begin{eqnarray}
\left( \psi h_{\kappa }h_{\bar{\mu}}h_{\nu }h_{\bar{\rho}}\psi \right) 
\mathrm{vol} &=&\psi _{abcd}\psi _{mnpq}h_{\kappa }^{am}h_{\bar{\mu}%
}^{bn}h_{\nu }^{cp}h_{\bar{\rho}}^{dq}\mathrm{vol}  \notag \\
&=&24\left\langle \psi _{abcd}h_{\kappa }^{a}\wedge h_{\bar{\mu}}^{b}\wedge
h_{\nu }^{c}\wedge h_{\bar{\rho}}^{d},\ast \psi \right\rangle \mathrm{vol} 
\notag \\
&=&24\psi _{abcd}h_{\kappa }^{a}\wedge h_{\bar{\mu}}^{b}\wedge h_{\nu
}^{c}\wedge h_{\bar{\rho}}^{d}\wedge \varphi   \label{psi4hyuk}
\end{eqnarray}%
Hence, can rewrite (\ref{Rkmnr}) as 
\begin{eqnarray}
\mathcal{R}_{\kappa \bar{\mu}\nu \bar{\rho}} &=&\frac{1}{3}\left( G_{\bar{\mu%
}\nu }G_{\kappa \bar{\rho}}+G_{\bar{\mu}\kappa }G_{\nu \bar{\rho}}\right)
-G^{\tau \bar{\sigma}}A_{\bar{\mu}\tau \bar{\rho}}A_{\kappa \nu \bar{\sigma}%
}-\frac{5}{21}C_{\bar{\mu}\bar{\rho}}C_{\kappa \nu } \\
&&+\frac{8}{9}\frac{1}{V}\int \psi _{abcd}h_{\kappa }^{a}\wedge h_{\bar{\mu}%
}^{b}\wedge h_{\nu }^{c}\wedge h_{\bar{\rho}}^{d}\wedge \varphi   \notag \\
&&+\frac{1}{81}\frac{1}{V}\int \left( 5\func{Tr}\left( h_{(\kappa }h_{\bar{%
\mu}}\right) \func{Tr}\left( h_{\nu }h_{\bar{\rho})}\right) -6\func{Tr}%
\left( h_{\kappa }h_{\bar{\mu}}h_{\nu }h_{\bar{\rho}}\right) \right) \mathrm{%
vol}  \notag
\end{eqnarray}%
\qquad 

Note that because in the $\Lambda _{27}^{3}$ directions the first derivative
of $V$ vanishes, some of these terms which appear in the curvature
expression can also be expressed as derivatives of \ $V$: 
\begin{eqnarray*}
\frac{\partial ^{3}V}{\partial \bar{z}^{\mu }\partial z^{\nu }\partial \bar{z%
}^{\rho }} &=&-\frac{1}{3}A_{\bar{\mu}\nu \bar{\rho}} \\
\frac{\partial ^{4}V}{\partial z^{\kappa }\partial \bar{z}^{\mu }\partial
z^{\nu }\partial \bar{z}^{\rho }} &=&-\frac{8}{27}\int \psi _{abcd}h_{\kappa
}^{a}\wedge h_{\bar{\mu}}^{b}\wedge h_{\nu }^{c}\wedge h_{\bar{\rho}%
}^{d}\wedge \varphi  \\
&&+\frac{1}{243}\int \left( 6\func{Tr}\left( h_{\kappa }h_{\bar{\mu}}h_{\nu
}h_{\bar{\rho}}\right) -5\func{Tr}\left( h_{(\kappa }h_{\bar{\mu}}\right) 
\func{Tr}\left( h_{\nu }h_{\bar{\rho})}\right) \right) \mathrm{vol}
\end{eqnarray*}%
So alternatively, can write 
\begin{eqnarray*}
\mathcal{R}_{\kappa \bar{\mu}\nu \bar{\rho}} &=&\frac{1}{3}\left( G_{\bar{\mu%
}\nu }G_{\kappa \bar{\rho}}+G_{\bar{\mu}\kappa }G_{\nu \bar{\rho}}\right)
-G^{\tau \bar{\sigma}}A_{\bar{\mu}\tau \bar{\rho}}A_{\kappa \nu \bar{\sigma}%
}-\frac{5}{21}C_{\bar{\mu}\bar{\rho}}C_{\kappa \nu } \\
&&-\frac{3}{V}\frac{\partial ^{4}V}{\partial z^{\kappa }\partial \bar{z}%
^{\mu }\partial z^{\nu }\partial \bar{z}^{\rho }}
\end{eqnarray*}%
Define 
\begin{equation}
U_{\bar{M}}=\frac{3}{V}\frac{\partial ^{3}V}{\partial \bar{z}^{M}\partial
z^{N}\partial \bar{z}^{R}}G^{N\bar{R}}  \label{UM}
\end{equation}%
Then, 
\begin{equation}
\partial _{K}U_{\bar{M}}=\frac{3}{V}\left( \frac{\partial ^{4}V}{\partial
z^{K}\partial \bar{z}^{M}\partial z^{N}\partial \bar{z}^{R}}G^{N\bar{R}}-%
\frac{\partial ^{3}V}{\partial \bar{z}^{M}\partial z^{N}\partial \bar{z}^{R}}%
A_{K}^{\ \ N\bar{R}}\right)   \label{dkum}
\end{equation}%
We can use this to express the Ricci curvature%
\begin{equation}
\mathcal{R}_{\kappa \bar{\mu}}=\left( \frac{1}{3}b^{3}\left( X\right) -\frac{%
1}{63}\right) G_{\kappa \bar{\mu}}-\partial _{\kappa }U_{\bar{\mu}}
\label{riccimn}
\end{equation}%
where $b^{3}\left( X\right) =b_{27}^{3}+1$ is the third Betti number of $X$.
Also, 
\begin{eqnarray}
\mathcal{R}_{0\bar{\mu}} &=&-aA_{\bar{\mu}\nu \bar{\rho}}G^{\nu \bar{\rho}%
}=-\partial _{0}U_{\bar{\mu}}  \label{ricci0m} \\
\mathcal{R}_{0\bar{0}} &=&2a^{2}b^{3}\left( X\right)   \label{ricci00}
\end{eqnarray}

Although here we have certain similarities with the structure of the
Calabi-Yau moduli space, but we are lacking a key feature of Calabi-Yau
moduli space - a particular line bundle over the moduli space. For example,
the holomorphic $3$-form on a Calabi-Yau $3$-fold defines a complex line
bundle over the complex structure moduli space. In the $G_{2}$ case, we
could try and see what happens if we look at the real line bundle $L$
defined by $\varphi $ over the complexified $G_{2}$ moduli space $\mathcal{M}%
_{\mathbb{C}}$. So consider the gauge transformations 
\begin{equation}
\varphi \longrightarrow f\left( \func{Re}z\right) \varphi
\label{gaugetransform}
\end{equation}%
where each $f\left( z\right) $ is a real number. Then, as in \cite%
{deBoer:2005pt}, define a covariant derivative $\mathcal{D}$ on $L$ by 
\begin{equation}
\mathcal{D}_{M}\varphi =\partial _{M}\varphi +\frac{1}{7}\left( \partial
_{M}K\right) \varphi .  \label{covderiv}
\end{equation}%
Under the transformation (\ref{gaugetransform}) 
\begin{eqnarray*}
V &\longrightarrow &f^{\frac{7}{3}}V \\
K &\longrightarrow &K-7\log f
\end{eqnarray*}%
and so 
\begin{equation*}
\partial _{M}K\longrightarrow \partial _{M}K-\frac{7}{f}\partial _{M}f.
\end{equation*}%
Hence, 
\begin{equation}
\mathcal{D}_{M}\varphi \longrightarrow f\mathcal{D}_{M}\varphi .
\label{dmphitrans}
\end{equation}%
Moreover, from the expression for $\partial _{M}K$ (\ref{dkdsm}), we find
that 
\begin{equation*}
\mathcal{D}_{0}\varphi =0\ \ \ \ \ \mathcal{D}_{\mu }\varphi =\partial _{\mu
}\varphi
\end{equation*}%
\qquad \qquad

So as noted in \cite{deBoer:2005pt}, this covariant derivative projects out
the $G_{2}$ singlet contribution. It also gives a covariant way in which to
extract the $\mathbf{27}$ contributions so we can use $\mathcal{D}%
_{M}\varphi $ when just need to extract $\partial _{\mu }\varphi $. Also
consider%
\begin{eqnarray}
\frac{1}{V}\left\langle \left\langle \mathcal{D}_{M}\varphi ,\ast \mathcal{D}%
_{\bar{N}}\varphi \right\rangle \right\rangle &=&\frac{1}{V}\int \mathcal{D}%
_{M}\varphi \wedge \ast \mathcal{D}_{\bar{N}}\varphi  \label{27metric} \\
&=&G_{M\bar{N}}-\frac{1}{7}\partial _{M}K\partial _{\bar{N}}K.  \notag
\end{eqnarray}%
When one of the indices is equal to zero, the whole expression vanishes.
However if both refer to the $27$-dimensional components, then we just get $%
G_{\mu \bar{\nu}}$. A similar expression holds for $C_{MN}$.

More generally, we can extend the covariant to any quantity which transforms
under (\ref{gaugetransform}). Suppose $Q\left( z\right) $ is a function on $%
\mathcal{M}_{\mathbb{C}}$, which under (\ref{gaugetransform}) transforms as 
\begin{equation*}
Q\left( z\right) \longrightarrow f\left( z\right) ^{a}Q\left( z\right) .
\end{equation*}%
Then define the covariant derivative on it by 
\begin{equation}
\mathcal{D}_{M}Q=\partial _{M}Q+\frac{a}{7}\left( \partial _{M}K\right) Q.
\label{dmq}
\end{equation}%
From this we get 
\begin{eqnarray*}
\mathcal{D}_{M}V &=&0 \\
\mathcal{D}_{M}\left( \ast \varphi \right) &=&\partial _{M}\left( \ast
\varphi \right) +\frac{1}{7}\frac{4}{3}\left( \partial _{M}K\right) \left(
\ast \varphi \right)
\end{eqnarray*}%
and in particular, 
\begin{equation*}
\mathcal{D}_{0}\left( \ast \varphi \right) =0\ \ \ \ \ \mathcal{D}_{\mu
}\left( \ast \varphi \right) =-\ast \left( \partial _{\mu }\varphi \right)
\end{equation*}%
so, in fact 
\begin{equation*}
\mathcal{D}_{M}\left( \ast \varphi \right) =-\ast \mathcal{D}_{M}\varphi .
\end{equation*}%
\qquad

Further we can extend $\mathcal{D}_{M}$ to objects with moduli space indices
by replacing $\partial $ by $\nabla $ - the metric-compatible covariant
derivative with respect to the moduli space metric $G_{M\bar{N}}$. For which
the Christoffel symbols are given by 
\begin{equation}
\Gamma _{M\ Q}^{\ \ N}=G^{N\bar{P}}\partial _{M}G_{\bar{P}Q}=A_{\ MQ}^{N}
\label{christoffel}
\end{equation}%
With these Christoffel symbols the covariant derivative of $C_{MN}$ is hence 
\begin{equation}
\nabla _{Q}C_{MN}=-A_{QMN}.  \label{dqcmn}
\end{equation}%
Then we also find that 
\begin{eqnarray}
\mathcal{D}_{M}\mathcal{D}_{N}\varphi &=&\partial _{M}\left( \partial
_{N}\varphi +\frac{1}{7}\left( \partial _{N}K\right) \varphi \right) -A_{\
NM}^{P}\mathcal{D}_{P}\varphi +\frac{1}{7}\partial _{M}K\mathcal{D}%
_{N}\varphi  \notag \\
&=&\frac{1}{7}\left( C_{MN}-\frac{1}{7}\partial _{M}K\partial _{N}K\right)
\varphi -A_{\ NM}^{P}\mathcal{D}_{P}\varphi +\frac{2}{7}\partial _{(M}K%
\mathcal{D}_{N)}\varphi  \label{dmdnphi} \\
&=&\frac{1}{7}\frac{1}{V}\left\langle \left\langle \mathcal{D}_{M}\varphi
,\ast \mathcal{D}_{N}\varphi \right\rangle \right\rangle \varphi -A_{\
NM}^{P}\mathcal{D}_{P}\varphi +\frac{2}{7}\partial _{(M}K\mathcal{D}%
_{N)}\varphi
\end{eqnarray}%
and for mixed type derivatives, we have 
\begin{eqnarray*}
\mathcal{D}_{\bar{M}}\mathcal{D}_{N}\varphi &=&\partial _{\bar{M}}\left(
\partial _{N}\varphi +\frac{1}{7}\left( \partial _{N}K\right) \varphi
\right) +\frac{2}{7}\partial _{(\bar{M}}K\mathcal{D}_{N)}\varphi \\
&=&\frac{1}{7}\frac{1}{V}\left\langle \left\langle \mathcal{D}_{\bar{M}%
}\varphi ,\ast \mathcal{D}_{N}\varphi \right\rangle \right\rangle \varphi +%
\frac{2}{7}\partial _{(\bar{M}}K\mathcal{D}_{N})\varphi \\
&=&\frac{1}{7}\left( G_{\bar{M}N}\varphi +\frac{2}{7}\left( \partial _{\bar{M%
}}K\partial _{N}K\right) \varphi +\partial _{(\bar{M}}K\partial _{N)}\varphi
\right)
\end{eqnarray*}%
Note that here the covariant derivatives commute, so this connection is in
fact flat.

Now look at the third covariant derivative of $\varphi $ 
\begin{eqnarray}
\left\langle \left\langle \mathcal{D}_{R}\mathcal{D}_{M}\mathcal{D}%
_{N}\varphi ,\ast \varphi \right\rangle \right\rangle &=&\mathcal{D}%
_{R}\left\langle \left\langle \mathcal{D}_{M}\mathcal{D}_{N}\varphi ,\ast
\varphi \right\rangle \right\rangle -\left\langle \left\langle \mathcal{D}%
_{M}\mathcal{D}_{N}\varphi ,\mathcal{D}_{R}\ast \varphi \right\rangle
\right\rangle  \notag \\
&=&\mathcal{D}_{R}\left\langle \left\langle \mathcal{D}_{M}\varphi ,\ast 
\mathcal{D}_{N}\varphi \right\rangle \right\rangle +\left\langle
\left\langle \mathcal{D}_{M}\mathcal{D}_{N}\varphi ,\ast \mathcal{D}%
_{R}\varphi \right\rangle \right\rangle  \label{3covphi1}
\end{eqnarray}%
First look at the second term in (\ref{3covphi1}). Since $\mathcal{D}%
_{R}\varphi \in \Lambda _{27}^{3}$, we basically get the projection $\pi
_{27}\left( \mathcal{D}_{M}\mathcal{D}_{N}\varphi \right) $:%
\begin{eqnarray*}
\left\langle \left\langle \mathcal{D}_{M}\mathcal{D}_{N}\varphi ,\ast 
\mathcal{D}_{R}\varphi \right\rangle \right\rangle &=&-A_{\
NM}^{P}\left\langle \left\langle \mathcal{D}_{P}\varphi ,\ast \mathcal{D}%
_{R}\varphi \right\rangle \right\rangle +\frac{2}{7}\partial
_{(M}K\left\langle \left\langle \mathcal{D}_{N)}\varphi ,\ast \mathcal{D}%
_{R}\varphi \right\rangle \right\rangle \\
&=&-A_{MNR}+\frac{1}{7}A_{\ MN}^{P\ \ \ \ \ }\partial _{R}K\partial _{P}K+%
\frac{2}{7}C_{R(N}\partial _{M)}K-\frac{2}{49}\partial _{R}K\partial
_{M}K\partial _{N}K
\end{eqnarray*}%
In the first term of (\ref{3covphi1}), we have 
\begin{eqnarray*}
\mathcal{D}_{R}\left\langle \left\langle \mathcal{D}_{M}\varphi ,\ast 
\mathcal{D}_{N}\varphi \right\rangle \right\rangle &=&V\mathcal{D}_{R}\left( 
\frac{1}{V}\left\langle \left\langle \mathcal{D}_{M}\varphi ,\ast \mathcal{D}%
_{N}\varphi \right\rangle \right\rangle \right) \\
&=&V\nabla _{R}\left\langle \left\langle \mathcal{D}_{M}\varphi ,\ast 
\mathcal{D}_{N}\varphi \right\rangle \right\rangle \\
&=&V\left( \nabla _{R}C_{MN}-\frac{1}{7}\nabla _{R}\left( \partial
_{M}K\partial _{N}K\right) \right) \\
&=&V\left( -A_{RMN}-\frac{2}{7}C_{R(M}\partial _{N)}K+\frac{2}{7}A_{\
R(M}^{P}\partial _{N)}K\partial _{P}K\right)
\end{eqnarray*}%
Combining, we overall obtain 
\begin{equation}
\frac{1}{V}\left\langle \left\langle \mathcal{D}_{R}\mathcal{D}_{M}\mathcal{D%
}_{N}\varphi ,\ast \varphi \right\rangle \right\rangle =-2A_{RMN}-\frac{2}{49%
}\partial _{R}K\partial _{M}K\partial _{N}K+\frac{3}{7}A_{(MN}^{\ \ \ \ \ \
P}\partial _{R)}K\partial _{P}K  \label{3covphi2}
\end{equation}%
Decomposing this into components, we have 
\begin{eqnarray*}
\frac{1}{V}\left\langle \left\langle \mathcal{D}_{\rho }\mathcal{D}_{\mu }%
\mathcal{D}_{\nu }\varphi ,\ast \varphi \right\rangle \right\rangle
&=&-2A_{\rho \mu \nu } \\
\frac{1}{V}\left\langle \left\langle \mathcal{D}_{0}\mathcal{D}_{\mu }%
\mathcal{D}_{\nu }\varphi ,\ast \varphi \right\rangle \right\rangle
&=&2C_{\mu \nu } \\
\frac{1}{V}\left\langle \left\langle \mathcal{D}_{0}\mathcal{D}_{0}\mathcal{D%
}_{\nu }\varphi ,\ast \varphi \right\rangle \right\rangle &=&0 \\
\frac{1}{V}\left\langle \left\langle \mathcal{D}_{0}\mathcal{D}_{0}\mathcal{D%
}_{0}\varphi ,\ast \varphi \right\rangle \right\rangle &=&0
\end{eqnarray*}%
Therefore, the quantity$\frac{1}{V}\left\langle \left\langle \mathcal{D}%
_{\rho }\mathcal{D}_{\mu }\mathcal{D}_{\nu }\varphi ,\ast \varphi
\right\rangle \right\rangle $ essentially gives the Yukawa coupling, again
giving a result analogous to the case of Calabi-Yau moduli spaces.

\section{Concluding remarks}

In this paper, we have computed the curvature of the complexified $G_{2}$
moduli space and found that while it has terms which are similar to the
curvature of Calabi-Yau moduli, there are a number of new terms. In future
work it would be interesting to interpret these new terms geometrically. If
we consider a $7$-manifold of the form $CY_{3}\times S^{1}$ where $CY_{3}$
is a Calabi-Yau $3$-fold, then we can define a torsion-free $G_{2}$
structure on it. The relationship between the Calabi-Yau moduli space and
the $G_{2}$ moduli space is however very non-trivial, because the complex
structure moduli and the K\"{a}hler structure moduli become intertwined with
each other. So it could turn out to be illuminating to try and relate the
curvature of the $G_{2}$ moduli space to the curvatures of complex and K\"{a}%
hler moduli spaces. In that case, however, $b_{7}^{3}=1$, so in fact the
second derivative of our K\"{a}hler potential would give a pseudo-K\"{a}hler
metric with signature $\left( -+...+\right) $ (\ref{d2kss}). Moreover, the
ansatz for the $C$-field (\ref{Cansatz}) would also have to be different.
Understanding how the Calabi-Yau moduli space is related to the $G_{2}$
moduli space could also enable us to find a manifestation of mirror symmetry
from the $G_{2}$ perspective. Moreover, it would be interesting to see how
existing approaches to mirror symmetry on $G_{2}$ manifolds (such as \cite%
{Gukov:2002jv}) affect the geometric structures on the moduli space.

Another possible direction for further research is to look at $G_{2}$
manifolds in a slightly different way. Suppose we have type $IIA$
superstrings on a non-compact Calabi-Yau $3$-fold with a special Lagrangian
submanifold which is wrapped by a $D6$ brane which also fills $M_{4}$. Then,
as explained in \cite{VafaKlemm:2001nx}, from the $M$-theory perspective
this looks like a $S^{1}$ bundle over the Calabi-Yau which is degenerate
over the special Lagrangian submanifold, but this $7$-manifold is still a $%
G_{2}$ manifold. The moduli space of this manifold will be then determined
by the Calabi-Yau moduli and the special Lagrangian moduli. This possibly
could provide more information about mirror symmetry on Calabi-Yau manifolds 
\cite{StromingerYau:1996it}.

\appendix

\section{Appendix A: Projections of $3$-forms}

\setcounter{equation}{0}Here will prove the formulae (\ref{p1chi}) to (\ref%
{p27chi}) which give the projections of $3$-forms into $1$-dimensional, $7$%
-dimensional and $27$-dimensional components. Let $\chi \in \Lambda ^{3}$.
Since $\Lambda _{1}^{3},\Lambda _{7}^{3}$ and $\Lambda _{27}^{3}$ are all
orthogonal to each other, we immediately get 
\begin{equation*}
\pi _{1}\left( \chi \right) =a\varphi \ \text{where }a=\frac{1}{42}\left(
\chi _{abc}\varphi ^{abc}\right) =\frac{1}{7}\,\left\langle \chi ,\varphi
\right\rangle \ \text{and }\left\vert \pi _{1}\left( \chi \right)
\right\vert ^{2}=7a^{2}\text{.}
\end{equation*}%
To work out $\pi _{7}\left( \chi \right) $, suppose 
\begin{equation*}
\pi _{7}\left( \chi \right) =u\lrcorner \ast \varphi
\end{equation*}%
then consider 
\begin{eqnarray}
\left( u\lrcorner \ast \varphi \right) \wedge \ast \left( v\lrcorner \ast
\varphi \right) &=&\left( u\lrcorner \ast \varphi \right) \wedge \varphi
\wedge v^{\flat }  \notag \\
&=&4\ast u^{\flat }\wedge v^{\flat }=4\left\langle u,v\right\rangle \mathrm{%
vol}  \label{p7calc1}
\end{eqnarray}%
So this gives 
\begin{equation}
\left\vert \pi _{7}\left( \chi \right) \right\vert ^{2}=4\left\vert \omega
\right\vert ^{2}  \label{p7norm}
\end{equation}%
However (\ref{p7calc1}) can also be expressed as 
\begin{eqnarray}
\left( u\lrcorner \ast \varphi \right) \wedge \ast \left( v\lrcorner \ast
\varphi \right) &=&\frac{1}{6}\pi _{7}\left( \chi \right) _{mnp}v_{a}\psi
^{amnp}\mathrm{vol}  \notag \\
&=&-\frac{1}{6}\pi _{7}\left( \chi \right) _{mnp}\psi ^{mnpa}v_{a}\mathrm{vol%
}  \label{p7calc2}
\end{eqnarray}%
Equating (\ref{p7calc1}) and (\ref{p7calc2}), we get 
\begin{equation*}
u^{a}=-\frac{1}{24}\pi _{7}\left( \chi \right) _{mnp}\psi ^{mnpa}=\omega ^{a}%
\text{.}
\end{equation*}%
Finally we look at $\pi _{27}\left( \chi \right) $. Consider 
\begin{equation*}
\chi _{abc}=\pi _{1}\left( \chi \right) _{abc}+\pi _{7}\left( \chi \right)
_{abc}+h_{[a}^{d}\varphi _{bc]d}
\end{equation*}%
Then, 
\begin{eqnarray}
\pi _{1}\left( \chi \right) _{mn\{a}\varphi _{b\}}^{\ \ mn} &=&a\varphi
_{mn\{a}\varphi _{b\}}^{\ mn}=6g_{\{ab\}}=0  \label{p27p1} \\
\pi _{7}\left( \chi \right) _{mn\{a}\varphi _{b\}}^{\ \ mn} &=&\omega
^{p}\psi _{pmn\{a}\varphi _{b\}}^{\ mn}=4v^{p}\varphi _{p\{ab\}}=0
\label{p7p1}
\end{eqnarray}%
Therefore, 
\begin{eqnarray}
\frac{3}{4}\chi _{mn\{a}\varphi _{b\}}^{\ \ mn} &=&\frac{3}{4}%
h_{[m}^{d}\varphi _{n\{a]d}\varphi _{b\}}^{\ mn}  \notag \\
&=&\frac{1}{2}h_{m}^{d}\varphi _{n\{a\left\vert d\right\vert }\varphi
_{b\}}^{\ mn}+\frac{1}{4}\varphi _{mnd}h_{\{a}^{d}\varphi _{b\}}^{\ mn} 
\notag \\
&=&\frac{1}{2}h_{m}^{d}\left( g_{\{ab\}}\delta _{d}^{m}-\delta
_{\{a}^{m}g_{b\}d}-\psi _{\{ab\}d}^{\ \ \ \ \ \ \ \ m}\right) +\frac{3}{2}%
h_{ab}  \notag \\
&=&h_{ab}  \label{p27hab}
\end{eqnarray}%
as required. Moreover, 
\begin{eqnarray}
\left\vert \pi _{27}\left( \chi \right) \right\vert ^{2} &=&\frac{1}{6}%
h_{[a}^{d}\varphi _{bc]d}h^{ea}\varphi _{\ \ \ e}^{bc}  \notag \\
&=&\frac{1}{18}h_{a}^{d}\varphi _{bcd}h^{ea}\varphi _{\ \ \ e}^{bc}+\frac{1}{%
9}h_{c}^{d}\varphi _{abd}h^{ea}\varphi _{\ \ \ e}^{bc}  \notag \\
&=&\frac{1}{3}\left\vert h\right\vert ^{2}-\frac{1}{9}h_{c}^{d}h^{ea}\left(
\delta _{a}^{c}g_{de}-g_{ae}\delta _{q}^{c}+\ast \varphi _{\ ade}^{c}\right)
\notag \\
&=&\frac{2}{9}\left\vert h\right\vert ^{2}  \label{p27normh}
\end{eqnarray}

\section{Appendix B: Determinants}

\setcounter{equation}{0}In this section, we will review deformations of
determinants. Let $I$ be the $n\times n$ identity matrix, and let $h$ be a
symmetric $n\times n$ matrix. Suppose $\lambda _{1},...,\lambda _{n}$ are
eigenvalues of $h$. Then 
\begin{eqnarray}
\det \left( I+\varepsilon h\right) &=&\prod_{i=1}^{n}\left( 1+\varepsilon
\lambda _{i}\right)  \label{det1ph} \\
&=&1+\varepsilon \sum_{i}\lambda _{i}+\varepsilon ^{2}\sum_{i<j}\lambda
_{i}\lambda _{j}+\varepsilon ^{3}\sum_{i<j<k}\lambda _{i}\lambda _{j}\lambda
_{k}+\varepsilon ^{4}\sum_{i<j<k<l}\lambda _{i}\lambda _{j}\lambda
_{k}\lambda _{l}+O\left( \varepsilon ^{5}\right)  \notag
\end{eqnarray}%
Define 
\begin{equation*}
t_{k}=\sum_{i}\lambda _{i}^{k}=\func{Tr}\left( h^{k}\right) .
\end{equation*}%
Then from Newton's identities we know that%
\begin{eqnarray*}
\sum_{i}\lambda _{i} &=&t_{1} \\
\sum_{i<j}\lambda _{i}\lambda _{j} &=&\frac{1}{2}\left(
t_{1}^{2}-t_{2}\right) \\
\sum_{i<j<k}\lambda _{i}\lambda _{j}\lambda _{k} &=&\frac{1}{6}\left(
t_{1}^{3}-3t_{1}t_{2}+2t_{3}\right) \\
\sum_{i<j<k<l}\lambda _{i}\lambda _{j}\lambda _{k}\lambda _{l} &=&\frac{1}{24%
}\left( t_{1}^{4}-6t_{1}^{2}t_{2}+3t_{2}^{2}+8t_{1}t_{3}-6t_{4}\right)
\end{eqnarray*}%
and so we obtain 
\begin{eqnarray}
\det \left( I+\varepsilon h\right) &=&1+\varepsilon t_{1}+\frac{1}{2}%
\varepsilon ^{2}\left( t_{1}^{2}-t_{2}\right) +\frac{1}{6}\varepsilon
^{3}\left( t_{1}^{3}-3t_{1}t_{2}+2t_{3}\right)  \label{detIph2} \\
&&+\frac{1}{24}\varepsilon ^{4}\left(
t_{1}^{4}-6t_{1}^{2}t_{2}+3t_{2}^{2}+8t_{1}t_{3}-6t_{4}\right) +O\left(
\varepsilon ^{5}\right) .  \notag
\end{eqnarray}%
Now, for a metric $g$, we get 
\begin{eqnarray*}
\frac{\det \left( g+\varepsilon h\right) }{\det g} &=&1+\varepsilon t_{1}+%
\frac{1}{2}\varepsilon ^{2}\left( t_{1}^{2}-t_{2}\right) +\frac{1}{6}%
\varepsilon ^{3}\left( t_{1}^{3}-3t_{1}t_{2}+2t_{3}\right) \\
&&+\frac{1}{24}\varepsilon ^{4}\left(
t_{1}^{4}-6t_{1}^{2}t_{2}+3t_{2}^{2}+8t_{1}t_{3}-6t_{4}\right) +O\left(
\varepsilon ^{5}\right)
\end{eqnarray*}%
where the traces are now with respect to the metric $g$.

\bibliographystyle{jhep2}
\bibliography{refs2}

\providecommand{\href}[2]{#2}\begingroup\raggedright\begin{thebibliography}{10}

\bibitem{Candelas:1990pi}
P.~Candelas and X.~de~la Ossa, {\it Moduli space of {C}alabi-{Y}au manifolds},
  {\em Nucl. Phys.} {\bf B355} (1991) 455--481.

\bibitem{Strominger:1990pd}
A.~Strominger, {\it Special geometry},  {\em Commun. Math. Phys.} {\bf 133}
  (1990) 163--180.

\bibitem{CalabiYau}
S.-T. Yau, {\it On the {R}icci curvature of a compact {K}aehler manifold and
  the complex monge-amp\`ere equation. {I}},  {\em Comm. Pure Appl. Math.} {\bf
  31} (1978) 339--411.

\bibitem{Joycebook}
D.~D. Joyce, {\em Compact manifolds with special holonomy}.
\newblock Oxford Mathematical Monographs. Oxford University Press, 2000.

\bibitem{Kovalev:2001zr}
A.~Kovalev, {\it Twisted connected sums and special {R}iemannian holonomy},
  \href{http://arXiv.org/abs/math/0012189}{{\tt math/0012189}}.

\bibitem{Gibbons:1989er}
G.~W. Gibbons, D.~N. Page and C.~N. Pope, {\it Einstein metrics on ${S}^{3}$,
  ${R}^{3}$ and ${R}^{3}$ bundles},  {\em Commun. Math. Phys.} {\bf 127} (1990)
  529.

\bibitem{bryant-2003}
R.~L. Bryant, {\it Some remarks on {G}\_2-structures},
  \href{http://arXiv.org/abs/math/0305124}{{\tt math/0305124}}.

\bibitem{deBoer:2005pt}
J.~de~Boer, A.~Naqvi and A.~Shomer, {\it The topological {G}(2) string},
  \href{http://arXiv.org/abs/hep-th/0506211}{{\tt hep-th/0506211}}.

\bibitem{karigiannis-2007a}
S.~Karigiannis and N.~C. Leung, {\it Hodge theory for {G}2-manifolds:
  Intermediate {J}acobians and {A}bel-{J}acobi maps},
  \href{http://arXiv.org/abs/0709.2987}{{\tt 0709.2987}}.

\bibitem{karigiannis-2005-57}
S.~Karigiannis, {\it Deformations of {G}\_2 and {S}pin(7) {S}tructures on
  {M}anifolds},  {\em Canadian Journal of Mathematics} {\bf 57} (2005) 1012
  [\href{http://arXiv.org/abs/math/0301218}{{\tt math/0301218}}].

\bibitem{FernandezGray}
M.~Fern{\'a}ndez and A.~Gray, {\it Riemannian manifolds with structure group
  {$G\sb{2}$}},  {\em Ann. Mat. Pura Appl. (4)} {\bf 132} (1982) 19--45 (1983).

\bibitem{Berger1955}
M.~Berger, {\it Sur les groupes d'holonomie homog\`ene des vari\'et\'es \`a
  connexion affine et des vari\'et\'es riemanniennes},  {\em Bull. Soc. Math.
  France} {\bf 83} (1955).

\bibitem{karigiannis-2007}
S.~Karigiannis, {\it Geometric {F}lows on {M}anifolds with ${G}\_2$
  {S}tructure, {I}},  \href{http://arXiv.org/abs/math/0702077}{{\tt
  math/0702077}}.

\bibitem{House:2004pm}
T.~House and A.~Micu, {\it M-theory compactifications on manifolds with {G}(2)
  structure},  {\em Class. Quant. Grav.} {\bf 22} (2005) 1709--1738
  [\href{http://arXiv.org/abs/hep-th/0412006}{{\tt hep-th/0412006}}].

\bibitem{AcharyaGukov}
B.~S. Acharya and S.~Gukov, {\it M theory and {S}ingularities of {E}xceptional
  {H}olonomy {M}anifolds},  {\em Phys. Rept.} {\bf 392} (2004) 121--189
  [\href{http://arXiv.org/abs/hep-th/0409191}{{\tt hep-th/0409191}}].

\bibitem{Cremmer:1978km}
E.~Cremmer, B.~Julia and J.~Scherk, {\it Supergravity theory in 11 dimensions},
   {\em Phys. Lett.} {\bf B 76} (1978) 409.

\bibitem{Witten:1996md}
E.~Witten, {\it On flux quantization in {M}-theory and the effective action},
  {\em J. Geom. Phys.} {\bf 22} (1997) 1--13
  [\href{http://arXiv.org/abs/hep-th/9609122}{{\tt hep-th/9609122}}].

\bibitem{Harvey:1999as}
J.~A. Harvey and G.~W. Moore, {\it Superpotentials and membrane instantons},
  \href{http://arXiv.org/abs/hep-th/9907026}{{\tt hep-th/9907026}}.

\bibitem{WittenBeasley}
C.~Beasley and E.~Witten, {\it A note on fluxes and superpotentials in
  {M}-theory compactifications on manifolds of {G}(2) holonomy},  {\em JHEP}
  {\bf 07} (2002) 046 [\href{http://arXiv.org/abs/hep-th/0203061}{{\tt
  hep-th/0203061}}].

\bibitem{Gutowski:2001fm}
J.~Gutowski and G.~Papadopoulos, {\it Moduli spaces and brane solitons for {M}
  theory compactifications on holonomy {G}(2) manifolds},  {\em Nucl. Phys.}
  {\bf B615} (2001) 237--265 [\href{http://arXiv.org/abs/hep-th/0104105}{{\tt
  hep-th/0104105}}].

\bibitem{Papadopoulos:1995da}
G.~Papadopoulos and P.~K. Townsend, {\it Compactification of d = 11
  supergravity on spaces of exceptional holonomy},  {\em Phys. Lett.} {\bf
  B357} (1995) 300--306 [\href{http://arXiv.org/abs/hep-th/9506150}{{\tt
  hep-th/9506150}}].

\bibitem{Riegeom}
R.~Portugal, {\it The {R}iegeom package: abstract tensor calculation},  {\em
  Comput. Phys. Commun.} {\bf 126} (2000) 261--268.

\bibitem{Hitchin:2000jd}
N.~J. Hitchin, {\it The geometry of three-forms in six and seven dimensions},
  \href{http://arXiv.org/abs/math/0010054}{{\tt math/0010054}}.

\bibitem{KarigiannisLin}
S.~Karigiannis and C.~Lin, {\it Curvature of the moduli space of
  ${G}_2$-manifolds},  {\em in preparation}.

\bibitem{Lee:2002fa}
J.-H. Lee and N.~C. Leung, {\it Geometric structures on {G}(2) and
  {S}pin(7)-manifolds},  \href{http://arXiv.org/abs/math/0202045}{{\tt
  math/0202045}}.

\bibitem{Gukov:2002jv}
S.~Gukov, S.-T. Yau and E.~Zaslow, {\it Duality and fibrations on {G}(2)
  manifolds},  \href{http://arXiv.org/abs/hep-th/0203217}{{\tt
  hep-th/0203217}}.

\bibitem{VafaKlemm:2001nx}
M.~Aganagic, A.~Klemm and C.~Vafa, {\it Disk instantons, mirror symmetry and
  the duality web},  {\em Z. Naturforsch.} {\bf A57} (2002) 1--28
  [\href{http://arXiv.org/abs/hep-th/0105045}{{\tt hep-th/0105045}}].

\bibitem{StromingerYau:1996it}
A.~Strominger, S.-T. Yau and E.~Zaslow, {\it Mirror symmetry is {T}-duality},
  {\em Nucl. Phys.} {\bf B479} (1996) 243--259
  [\href{http://arXiv.org/abs/hep-th/9606040}{{\tt hep-th/9606040}}].

\end{thebibliography}\endgroup

\end{document}